\documentclass[preprint2]{emulateapj}
\usepackage{graphicx}
\usepackage{txfonts}
\usepackage{url}
\graphicspath{{./fig/}{./png/}}
\newcommand{\EQ}{\begin{equation}}
\newcommand{\EN}{\end{equation}}
\newcommand{\EQA}{\begin{eqnarray}}
\newcommand{\ENA}{\end{eqnarray}}
\newcommand{\eq}[1]{(\ref{#1})}
\newcommand{\EEq}[1]{Equation~(\ref{#1})}
\newcommand{\Eq}[1]{Equation~(\ref{#1})}
\newcommand{\Eqs}[2]{Equations~(\ref{#1}) and~(\ref{#2})}

\newcommand{\App}[1]{Appendix~\ref{#1}}
\newcommand{\Sec}[1]{Section~\ref{#1}}
\newcommand{\Secs}[2]{Sections~\ref{#1} and \ref{#2}}
\newcommand{\Fig}[1]{Figure~\ref{#1}}
\newcommand{\FFig}[1]{Figure~\ref{#1}}
\newcommand{\Tab}[1]{Table~\ref{#1}}

\newcommand{\bra}[1]{\langle #1\rangle}

\newcommand{\tildeFFFF}{\tilde{\mbox{\boldmath ${\cal F}$}}{}}{}
\newcommand{\meanFFFF}{\overline{\mbox{\boldmath ${\cal F}$}}{}}{}
\newcommand{\meanemf}{\overline{\cal E} {}}

{}
{}
{}
\newcommand{\meanEMF}{\overline{\mbox{\boldmath ${\cal E}$}}{}}{}
{}
{}
{}
{}
{}
\newcommand{\meanBB}{\overline{\mbox{\boldmath $B$}}{}}{}
{}
\newcommand{\meanGG}{\overline{\mbox{\boldmath $G$}}{}}{}
\newcommand{\tildeGG}{\tilde{\mbox{\boldmath $G$}}{}}{}
\newcommand{\meanJJ}{\overline{\mbox{\boldmath $J$}}{}}{}
\newcommand{\meanUU}{\overline{\mbox{\boldmath $U$}}{}}{}
{}
{}

\newcommand{\meanB}{\overline{B}}

\newcommand{\meanC}{\overline{C}}
\newcommand{\meanG}{\overline{G}}

\newcommand{\meanJ}{\overline{J}}

\newcommand{\meanFFF}{\overline{\cal F}}
{}
{}
{}
{}
{}

\newcommand{\zzz}{\hat{\mbox{\boldmath $z$}} {}}

\newcommand{\kk}{\mbox{\boldmath $k$} {}}

\newcommand{\uu}{\mbox{\boldmath $u$} {}}
\newcommand{\UU}{\mbox{\boldmath $U$} {}}

\newcommand{\bb}{\mbox{\boldmath $b$} {}}
\newcommand{\BB}{\mbox{\boldmath $B$} {}}
\newcommand{\CC}{\mbox{\boldmath $C$} {}}

\newcommand{\ff}{\mbox{\boldmath $f$} {}}

\newcommand{\GG}{\mbox{\boldmath $G$} {}}

\newcommand{\nab}{\mbox{\boldmath $\nabla$} {}}

\newcommand{\SSSS}{\mbox{\boldmath ${\sf S}$} {}}

\newcommand{\ii}{{\rm i}}

\newcommand{\dd}{{\rm d} {}}

\def\St{\mbox{\rm St}}

\def\Rm{\mbox{\rm Re}_M}
\def\Rmc{\mbox{\rm Re}_{M,{\rm crit}}}
\def\Rey{\mbox{\rm Re}}
\def\Pe{\mbox{\rm Pe}}

\def\kf{k_{\rm f}}
\def\urms{u_{\rm rms}}
\def\kappat{\kappa_{\rm t}}
\def\kappatz{\kappa_{\rm t0}}
\def\etat{\eta_{\rm t}}
\def\etatz{\eta_{\rm t0}}

\def\half{{\textstyle{1\over2}}}

\def\onethird{{\textstyle{1\over3}}}

\newcommand{\yapj}[3]{ #1, {ApJ,} {#2}, #3}
\newcommand{\yptrsa}[3]{ #1, {Phil.\ Trans.\ R.\ Soc.\ A,} {#2}, #3}

\newcommand{\yan}[3]{ #1, {Astron.\ Nachr.,} {#2}, #3}

\newcommand{\yana}[3]{ #1, {A\&A,} {#2}, #3}

\newcommand{\ygafd}[3]{ #1, {Geophys.\ Astrophys.\ Fluid Dyn.,} {#2}, #3}

\newcommand{\yjfm}[3]{ #1, {J.\ Fluid Mech.,} {#2}, #3}

\newcommand{\ypf}[3]{ #1, {Phys.\ Fluids,} {#2}, #3}
\newcommand{\ypp}[3]{ #1, {Phys.\ Plasmas,} {#2}, #3}

\newcommand{\yprl}[3]{ #1, {Phys.\ Rev.\ Lett.,} {#2}, #3}

\newcommand{\ymn}[3]{ #1, {MNRAS,} {#2}, #3}
\newcommand{\ynat}[3]{ #1, {Nature,} {#2}, #3}

\newcommand{\ypre}[3]{ #1, {Phys.\ Rev.\ E,} {#2}, #3}

\newcommand{\yjour}[4]{ #1, {#2}, {#3}, #4}

\newcommand{\ybook}[3]{ #1, {#2} (#3)}

\begin{document}

\title{Memory effects in turbulent transport}
\author{Alexander Hubbard$^1$ and Axel Brandenburg$^{1,2}$}

\email{alex.i.hubbard@gmail.com
($ $Revision: 1.203 $ $)
}

\affil{
$^1$ NORDITA, AlbaNova University Center, Roslagstullsbacken 23,
SE 10691 Stockholm, Sweden\\
$^2$Department of Astronomy, AlbaNova University Center,
Stockholm University, SE 10691 Stockholm, Sweden
}

\begin{abstract}
In the mean-field theory of magnetic fields, turbulent transport,
i.e.\ the turbulent electromotive force, is
described by a combination of the $\alpha$ effect and turbulent magnetic diffusion,
 which are usually assumed to be
proportional respectively to the mean field and its spatial derivatives.
For a passive scalar there is just turbulent diffusion, where the mean
flux of concentration depends on the gradient of the mean concentration.
However, these proportionalities are approximations that are valid only if the
mean field or the mean concentration vary slowly in time.
Examples are presented where turbulent transport possesses memory, i.e.\ where it
depends crucially on the past history of the mean field.
Such effects are captured by replacing turbulent transport coefficients with time integral kernels,
resulting in transport coefficients that depend effectively on the frequency
or the growth rate of the mean field itself. In this paper we perform 
numerical experiments to find the characteristic timescale (or memory length)
of this effect as well as simple analytical
models of the integral kernels in the case of passive scalar
concentrations and kinematic dynamos.  The integral kernels can then be used to find
self-consistent growth or decay rates of the mean fields.
In mean-field dynamos the growth rates and cycle periods based on
steady state values of $\alpha$ effect and turbulent diffusivity
can be quite different from the actual values.
\end{abstract}


\section{Introduction}

A simple form of turbulent transport is the mixing of a passive
scalar associated with the mutual exchange of fluid parcels.
This process is similar to non-turbulent mixing that occurs just
because of thermal fluctuation or Brownian motion, often
referred to as molecular diffusion.
The latter process is described by a diffusion equation with a diffusion term of the
form $\kappa\nabla^2 C$, where $\kappa$ is the molecular diffusion
coefficient and $C$ is the concentration.
Turbulent diffusion, on the other hand, applies to a suitably
averaged mean concentration, $\meanC$, and is normally described by a
diffusion term of the form $\kappat\nabla^2\meanC$, where $\kappat$
is a turbulent diffusivity.
The ratio $\kappat/\kappa$ scales like the Reynolds number
(or, more precisely, the P\'eclet number) and can become very
large under many astrophysical conditions (stars, accretion discs, galaxies).

Problems connected with this simple prescription occur when the mean
concentration shows variations on timescales shorter than or
comparable to the correlation time of the turbulence.
In practice this means that a sinusoidal profile of $\meanC$ with
wavenumber $k$ would decay at a rate $\kappat k^2$ where $\kappat$
is no longer constant, but it depends itself on the actual decay rate.

The fact that problems occur when the mean concentration
changes on short timescales should not be surprising.
Indeed, in the text books of Moffatt (1978) and Krause \& R\"adler (1980)
it is shown that a proper description of turbulent transport involves
a convolution of an integral kernel with the mean concentration
over past times.
This is why one talks about memory effects: the turbulent
diffusion is not just an instantaneous property of the turbulence,
but depends on its full time history (Hori \& Yoshida 2008).
Dealing with a convolution over past times is an unpleasant complication,
so its effects are often neglected.
However, there can be circumstances of astrophysical relevance
where this is no longer permissible.

Such a circumstance is the damping of solar $p$-mode oscillations through
turbulent motions in the surface layers (Stix et al.\ 1993).
Here the timescales of $p$-modes and convection are comparable, so memory
effects must be important.
Stix et al.\ (1993) find that the turbulent diffusion is reduced by a
factor $\exp(-\omega_{\rm osc}\tau)$, where $\omega_{\rm osc}$ is the
oscillation frequency and $\tau$ is the correlation time of the turbulence.
Memory effects have also been invoked in connection with propagating front
solutions in the galactic dynamo (Fedotov et al.\ 2002, 2003), and variations
of the solar cycle (Otmianowska-Mazur et al.\ 1997), although there the
timescales are more disparate.

A practical way of dealing with memory effects has been proposed by
Blackman \& Field (2002, 2003), who derived an evolution equation
for the turbulent flux of concentration based on a simple closure
prescription known as the $\tau$ approximation.
One of the main beauties of this approach is that the usual diffusion
equation, which is of parabolic nature, is now replaced by a
damped wave equation, which is of hyperbolic nature.
This implies that signal propagation is no longer infinitely fast,
but its speed is limited to the rms velocity of the turbulence.
The principal validity of this approach has been demonstrated
using turbulence simulations of passive scalar diffusion
(Brandenburg et al.\ 2004).
One of the goals of the present paper is to provide a more direct means of
determining memory effects of turbulent transport that can also be applied
to more complicated cases of vector fields such as the magnetic field.

A promising method for calculating turbulent transport coefficients for the
magnetic field is the test-field method.
In this approach one calculates evolution equations for the small-scale
field that results from a given set of different test fields.
In this way one can calculate the full tensorial nature of the
turbulent diffusion tensor, as well as the $\alpha$ tensor that
can be relevant for amplifying the magnetic field if the turbulence
lacks mirror symmetry, for example in the presence of helicity.
These test fields have a given length scale characterized by some
wavenumber.
By varying this wavenumber it has been possible to determine the scale
dependence of the mean fields that are being diffused and/or amplified
(Brandenburg et al.\ 2008a).
Using a Fourier transformation over all wavenumbers, it is possible to
determine the spatial properties of the integral kernels that are used
in the convolution with the mean field over all other points in space.
It is customary to approximate the kernels by $\delta$ functions,
in which case the convolutions become multiplications.
In the test-field method, the corresponding coefficients are obtained
as the limit of vanishing wavenumber.
However, in order to make statements for finite domains of length $L$,
the wavenumber $k=k_1\equiv2\pi/L$ is most relevant.
Unless stated otherwise, we focus therefore on results for $k=k_1$.

In an analogous fashion, we can make the test fields time-dependent and
compute in this way the temporal properties of the integral kernels.
By imposing sinusoidal variations of the test fields over a range of
different frequencies we calculate the integral kernels first in Fourier
space, because there the convolution corresponds just to a multiplication.
The integral kernel in real space is then obtained by Fourier transformation.
Another possibility is to apply an exponentially growing or
decaying time variation.
In a sense this comes closest to the application of calculating modifications
of growth rates due to finite memory effects.
The integral kernel can then be calculated by inverse Laplace
transformation, but this approach involves integration along the imaginary
axis and is therefore only feasible if the data can be fitted reliably
to an analytic function.
We note that it is in principle also possible to determine integral kernels
directly by applying a $\delta$ function-like variation to the mean
concentration gradient or the mean field, but the
disadvantage here is that it is then not so easy to improve the statistics
by time averaging.
Nevertheless, such a $\delta$ function-like perturbation provides an
additional verification and is certainly a useful thought experiment.

The temporal properties of integral kernels in turbulent transport
may be particularly important in dynamo theory where simulations
and theory are now sufficiently accurate to show finite memory effects
under controlled conditions.
As a side effect, growth rates based on a dispersion relation with constant
$\alpha$ effect and turbulent magnetic diffusivity may become inaccurate.
It is quite plausible that under more complicated circumstances
finite memory effects will be even more important.
However, without proper knowledge of what to expect, this would only
remain speculation.
A goal of this paper is therefore to clarify finite memory effects
in simulations of forced helical turbulence in a periodic domain.
We consider here only the kinematic case, i.e.\ the velocity is
unaffected by the magnetic field.

In \Sec{Background}, we will motivate our work by considering two
approaches to calculating the growth rate of the Roberts flow dynamo.
In \Sec{formalism} we define our formalism,
most importantly the time response kernels that describe ``memory" effects.
We will treat both the turbulent transport of magnetic fields
and the conceptually simpler transport of passive scalars.
In \Sec{Preliminary} we give a brief theoretical overview before
discussing our numerical simulations and results in \Secs{sim}{sectionresults}.
We discuss those results in \Sec{Discussion} and conclude in \Sec{Conclusions}.

\section{Background: mismatch in growth rates}
\label{Background}

A direct approach to determining the growth rate of a dynamo is to solve
the induction equation for the magnetic field $\BB$ numerically:
\EQ
\frac{\partial\BB}{\partial t}=\nab\times(\UU\times\BB)+\eta\nabla^2\BB.
\label{dBdt}
\EN
Here $\UU$ is the velocity and $\eta$ is the microscopic magnetic diffusivity.
We are interested in dynamos that produce mean fields, $\meanBB$,
denoted here by an overbar.
In the following we take this to be an $xy$ average.
We calculate then the growth rate of the mean field as
\EQ
\lambda_{\rm growth}=\dd\ln\meanB_{\rm rms}/\dd t.
\EN
This can now be compared with the corresponding result from mean-field theory,
where one considers the averaged induction equation,
\EQ
\frac{\partial\meanBB}{\partial t}
=\nab\times(\meanUU\times\meanBB+\meanEMF)+\eta \nabla^2 \meanBB,
\label{meanfield}
\EN
with
\EQ
\meanEMF \equiv \overline{\uu \times \bb}
\EN
being the turbulent electromotive force and $\uu=\UU-\meanUU$ and $\bb=\BB-\meanBB$
are the fluctuations.
Symmetry considerations constrain the form of $\meanEMF$,
and in the case of homogeneous isotropic turbulence with helicity,
the expression for $\meanEMF$ is found to be
\EQ
\meanEMF=\alpha \meanBB-\etat\mu_0\meanJJ,
\label{ansatz}
\EN
where $\alpha$ describes the $\alpha$ effect, $\etat$ is the turbulent
magnetic diffusivity, $\meanJJ=\nab\times\meanBB/\mu_0$ is the mean
current density, $\mu_0$ is the vacuum permeability, and
higher order terms have been omitted.
Such a model is generally referred to as an $\alpha^2$ dynamo.
For references see Moffatt (1978) and Krause \& R\"adler (1980).

A new and accurate method for determining $\alpha$ and $\etat$
is the test-field method of Schrinner et al.\ (2005, 2007)
that will be described below.
The details of this method are not essential at this point,
except that we do emphasize that for our values of the magnetic Reynolds
number $\Rm$ the wavenumber
of the test field is chosen to be that of the box, which is also the
smallest wavenumber that fits into the domain.

For isotropic turbulence in a periodic domain the magnetic field can
develop long wavelength variations in any of the three coordinate
directions (Brandenburg 2001).  We assume this to
be the $z$ direction and use averages over the $x$ and $y$ directions.
Solutions of a homogeneous $\alpha^2$ dynamo
obey $\nab \times \meanBB=k_z\meanBB=\mu_0 \meanJJ$ and
are proportional to $\exp(\ii k_z z+\lambda t)$ with the dispersion
relation
\EQ \lambda=\alpha k_z-(\eta+\etat)k_z^2,
\label{disper}
\EN
where $k_z$ is the wavenumber in the $z$ direction.  Both $\alpha$
and $\etat$ are taken as constant in space owing to the assumed
statistical homogeneity of the turbulence.
For flows with positive kinetic helicity, $\alpha$ is expected to
be negative, so growing solutions correspond to negative values of $k_z$.

We refer to the value of $\lambda$ obtained from the dispersion relation
\eq{disper} as $\lambda_{\rm disp}$.
This is the second approach to determining the growth rate of the dynamo.
It has the disadvantage of being indirect, but the advantage of aiding
comprehension of the dynamo mechanism itself.
If the theory behind this second approach is correct, then the results
should match, so comparing the growth rates allows one to test the
validity of \Eq{ansatz}.

\begin{figure}[t!]
\centering\includegraphics[width=\columnwidth]{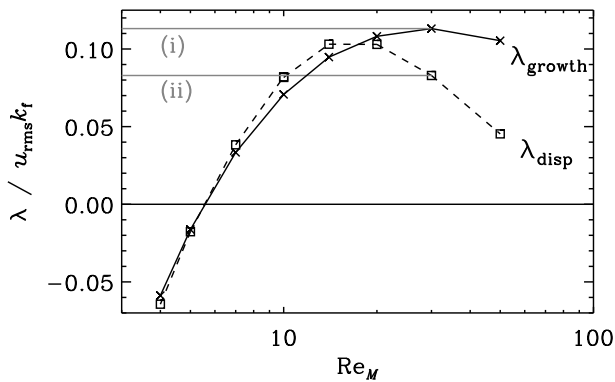}\caption{
$\Rm$ dependence of the growth rate for the Roberts flow as obtained
from a direct calculation ($\lambda_{\rm growth}$) compared with the
result of the dispersion relation,
$\lambda_{\rm disp}=\alpha k_z-(\eta+\etat)k_z^2$, using a
cubic domain of size $L^3$, where $k_1=2\pi/L$ and $\kf=\sqrt{2}k_1$.
For this range of $\Rm$, the most unstable mode is the largest one
that fits in the box ($k_z=k_1$).
The two horizontal lines in gray mark the values of $\lambda_{\rm growth}$
and $\lambda_{\rm disp}$ at $\Rm=30$, denoted by (i) and (ii), respectively.
}\label{pscan_Rm}\end{figure}

In order to motivate the purpose of this paper, let us now compare
in \Fig{pscan_Rm} $\lambda_{\rm growth}$ with $\lambda_{\rm disp}$ for
the simpler case of a steady periodic helical flow instead of turbulence.
We use here the Roberts flow, whose details will be discussed later.
The two estimates for $\lambda$ do indeed agree when $\lambda=0$,
at the critical magnetic Reynolds number for the onset of dynamo
action $\Rmc \simeq 5.52$.  For larger values of $\Rm$, there is a discrepancy
that can become rather dramatic for $\Rm>20$.
 
One of the motivations for our work then is the fact that the growth rate
estimated from \Eq{disper}, where $\alpha$ and $\etat$ are obtained
from the test-field method, becomes increasingly inaccurate for large
growth rates, implying that \Eq{ansatz} is inadequate to
describe growing dynamos.
We emphasize that this discrepancy vanishes not only in the marginal
case, but also for the nonlinearly saturated dynamo.
This is why in Brandenburg et al.\ (2008b) the quenched values of
$\alpha(\meanBB)$ and $\etat(\meanBB)$ were found to obey \Eq{disper}
with $\lambda=0$.

Even though the Roberts flow has been studied extensively over the years
(see, e.g., Roberts 1972, Soward 1987, Plunian et al.\ 1999,
Plunian \& R\"adler 2002a,b, Courvoisier 2008), and especially so in
connection with the Karlsruhe dynamo experiment
(cf.\ Stieglitz \& M\"uller 2001, R\"adler et al.\ 2002),
a discrepancy between theoretically expected growth rates based
on mean-field theory and the actual ones has never been reported.
For example in Plunian \& R\"adler (2002a), the actual growth rates
have been determined directly without invoking mean-field theory,
and in R\"adler et al.\ (2002) only the marginal case has been
compared with observations.
However, in the marginal case the discrepancy disappears.
In Plunian \& R\"adler (2002b), on the other hand,
the values of $\alpha$ and $\etat$ have again been determined
self-consistently for cases different from the marginal one.
Thus, the mean field is then of course no longer steady,
and so their values of $\alpha$ and $\etat$ apply only to
this particular time dependence, but not to a fictitious steady
case, for example.
We say here ``fictitious'', because for given values of $\Rm$ and $k_z$,
there is normally only one relevant solution, namely the one with the
largest value of $\lambda$.
However, for a predictive theory one should know $\alpha$ and $\etat$
before having solved the problem, i.e.\ before knowing $\lambda$.
In the following we explain how the fictitious steady case can actually be
realized in a simulation.

\begin{figure}[t!]
\centering\includegraphics[width=\columnwidth]{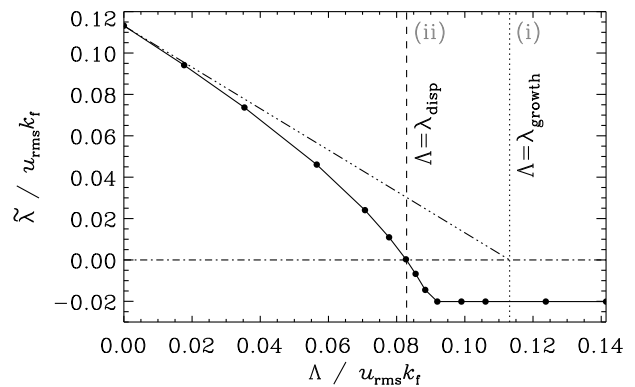}\caption{
Dependence of $\tilde\lambda$ on $\Lambda$ for $\Rm=30$.
The values of $\Lambda=\lambda_{\rm disp}$ and $\lambda_{\rm growth}$
of \Fig{pscan_Rm} are indicated by a vertical dashed
and dotted lines, denoted by (ii) and (i), respectively.
Note that $\tilde\lambda=0$ (dash-dotted line) for
$\Lambda=\lambda_{\rm disp}$, where 
$\lambda_{\rm disp}=\alpha k_z-(\eta+\etat)k_z^2$ with
$\alpha$ and $\etat$ obtained using the test-field method
for steady fields.
The linear interpolation between the points
$(\Lambda,\tilde\lambda)=(0,\lambda_{\rm growth})$ and
$(\lambda_{\rm growth},0)$ is indicated by a triple-dot-dash line.
}\label{pmean_friction}\end{figure}

In order to clarify the point that, for given values of $\Rm$,
$\alpha$ and $\etat$ depend also on the resulting growth rate,
let us now consider a modified induction equation
with an artificial ``friction'' term,
\EQ
\frac{\partial\BB}{\partial t}=\nab\times(\UU\times\BB)+\eta\nabla^2\BB
-\Lambda\meanBB,
\label{dBdtmod}
\EN
where $\Lambda$ is a new control parameter and $\meanBB$ is the
$xy$-averaged field.
Note that the evolution of the departure from this $xy$-averaged field,
$\bb=\BB-\meanBB$, is unaffected by this manipulation, so
$\meanEMF=\overline{\uu\times\bb}$ is exactly the same as before.
The solutions of $\BB$ have still an exponential time dependence,
and standard mean-field theory gives for the growth rate $\tilde\lambda$
of the mean field
\EQ
\tilde\lambda=\alpha k_z-(\eta+\etat)k_z^2-\Lambda.
\label{dispermod}
\EN
So, as the value of $\Lambda$ is increased (for given values of $\Rm$
and $k_z$), the growth rate $\tilde\lambda$ decreases.
[The tilde has been added to distinguish $\lambda$ from that used
in \Eq{disper}.]
There is a critical value $\Lambda_*$ for which $\tilde\lambda=0$.
This value is determined by
\EQ
\Lambda_*=\alpha k_z-(\eta+\etat)k_z^2.
\label{dispermod2}
\EN
Given that in this case the mean field is steady, we now expect
\Eq{dispermod2} to be accurate.
To verify this we solve \Eq{dBdtmod} numerically
and determine the growth rate $\tilde\lambda$.
The result is shown in \Fig{pmean_friction} where we plot
$\tilde\lambda$ vs.\ $\Lambda$ for $\Rm=30$.
For $\Lambda=0$ we find $\tilde\lambda=\lambda_{\rm growth}$.
More importantly, it turns out that $\tilde\lambda=0$ at a value
$\Lambda=\Lambda_*=\lambda_{\rm disp}$, indicated by (ii),
that is given by \Eq{dispermod2} with the same values of $\alpha$ and
$\etat$ that led earlier to the discrepancy in \Fig{pscan_Rm}.
Most crucially, the numerically determined growth rate
$\tilde\lambda$ in  \Fig{pmean_friction} deviates from
a linear interpolation between the points
$(\Lambda,\tilde\lambda)=(0,\lambda_{\rm growth})$ and
$(\lambda_{\rm growth},0)$.
This suggests again that the assumption of the $\alpha$ and $\etat$
in \Eq{dispermod} being independent of $\lambda$ is invalid.

We note that for larger values of $\Rm$ (e.g.\ for $\Rm=50$),
\Eq{dBdtmod} permits additional solutions with insignificant
$\meanBB$ that cannot be damped by the $\Lambda\meanBB$ term.
However, as a proof of concept, it was only essential that $\Rm$
was big enough so that there is a clear difference between
$\lambda_{\rm growth}$ and $\lambda_{\rm disp}$.

The results presented above show that a naive application of the
dispersion relation to cases where $\lambda\neq0$ is not possible
and gives results that disagree with the direct simulation.
This is because the values of $\alpha$ and $\etat$ apply only
to the steady case, as demonstrated by considering the
associated steady problem of \Eq{dBdtmod}, where $\Lambda=\Lambda_*$
is predicted from \Eq{dispermod2} using the $\alpha$ and $\etat$ values
obtained from the test-field method.

Recently, Hori \& Yoshida (2008) noted that, in the Roberts flow, memory
effects can be responsible for an enhancement of the growth rate.
The reason why Plunian \& R\"adler (2002b) found the correct growth rates
from \Eq{disper} even when $\lambda\neq0$ is that their values of
$\alpha$ and $\etat$ were automatically ``tuned'' to the resulting
growth rate.
Their values do therefore not apply to the steady case, which can
be verified by considering the mean-field problem associated with
\Eq{dBdtmod}.

To understand the reason for the discrepancy between actual
growth rates and those obtained from the standard (time-independent)
test-field method, it is important to recall that a multiplicative
relation in \Eq{ansatz} is only an approximation and that it should
instead be a convolution in space and time
(Moffatt 1978; Krause \& R\"adler 1980).
Alternatively, a Taylor series expansion of $\meanEMF$
in space and time can be employed.
Already in the simple case of the Roberts flow \Eq{ansatz}
cannot be justified when the mean field changes sufficiently rapidly
in time.
In this paper, we show that in such cases ``memory" effects of the turbulent
transport coefficients cannot be ignored.
This implies that the electromotive force at a given time depends not only on
the mean fields at that specific time, but also on the mean fields at all
prior times.
In practice, this means that the turbulent transport coefficients depend
themselves on the resulting growth rate and/or frequency of the
mean fields.

\section{Formalism}
\label{formalism}

Quite generally, we are interested in expressing quadratic correlations
of fluctuating quantities in terms of mean fields.
Examples include the mean turbulent concentration flux
and the mean turbulent electromotive force,
\EQ
\meanFFFF=\overline{\uu c},\quad\mbox{and}\quad
\meanEMF=\overline{\uu\times\bb},
\EN
respectively.
Here, $c=C-\meanC$ is the fluctuation of the concentration density.
The number of preferred directions available to mean quantities such
as $\meanFFFF$ and $\meanEMF$ are limited, and so
the aim is to relate them respectively
to the gradient of the mean concentration, $\meanGG=\nab\meanC$,
and to a linear combination of the mean magnetic field $\meanBB$ and its curl,
$\nab\times\meanBB=\mu_0\meanJJ$.
However, instead of multiplicative (instantaneous) relations of the form
\EQ
\meanFFFF=-\kappat\meanGG,\quad
\meanEMF=\alpha\meanBB-\etat\mu_0\meanJJ,
\label{FickianFE}
\EN
we now adopt such relations in their more general forms involving a
convolution in time, i.e.\
\EQ
\meanFFFF(t)=-\int_{-\infty}^t\hat\kappat(t-t')\meanGG(t')\,\dd t',
\label{convolutiont1}
\EN
and
\EQ
\meanEMF(t)=\int_{-\infty}^{t}\hat\alpha(t-t')\meanBB(t')\,\dd t'
-\int_{-\infty}^{t}\hat\etat(t-t')\mu_0\meanJJ(t')\,\dd t',
\label{meanEMF}
\EN
where quantities with a hat denote integral kernels, so
$\hat\kappat(t)$ is an integral kernel describing turbulent passive
scalar diffusion, $\hat\alpha(t)$ describes the $\alpha$ effect, and
$\hat\etat(t)$ the turbulent magnetic diffusion.
This approach is the most general search for memory effects, 
and we adopt it to find out how to modify \Eq{ansatz} to model more accurately
growing dynamos.

We recall that in general our averages (being two dimensional
over the $xy$ plane) are also functions of $z$, but the
$z$ dependence has here been suppressed in favor of a more compact notation.
In general, \Eqs{convolutiont1}{meanEMF} should also include a convolution over $z$.
This property has recently been studied in Brandenburg et al.\ (2008a), but
the spatial aspects of the convolution will here be ignored by considering
magnetic fields that have only a single wavenumber $k_z$, which corresponds
to the smallest wavenumber $k_1=2\pi/L$ that fits into the domain of
size $L^3$.

\subsection{Standard test-field methods}
\label{Standard}

In this section we reiterate the essence of the standard test-field
methods for calculating $\alpha$, $\etat$, and $\kappat$, where memory
effects are ignored.
As noted above, mean-field theory treats turbulent transport through the
correlations of fluctuating quantities as in \Eq{FickianFE}.  If the transported
quantity does not itself affect the dynamics of the system, as in the cases
of passive scalars or kinematic dynamos (where the magnetic field is too
weak to affect the momentum equation), then the transport coefficients
are functions of the velocity fields alone.

This independence of the transport coefficients of the mean field implies
that the transport coefficients will be found also in systems where a
mean field is externally imposed and does not obey any evolution equation.
Such a field is called a test field.
A set of different test fields is needed to determine simultaneously the
prefactors $\alpha$ and $\etat$ of $\meanBB$ and $\meanJJ$, respectively.
In the test-field method of Schrinner et al.\ (2005, 2007), one subtracts
the mean-field equation \eq{meanfield} from the full induction equation
\eq{dBdt} to obtain an evolution equation for the fluctuating field $\bb$,
\EQ
{\partial\bb\over\partial t}=\nab\times(\meanUU\times\bb+\uu\times\meanBB
+\uu\times\bb-\overline{\uu\times\bb})+\eta\nabla^2\bb.
\label{Testfield_b1}
\EN
This equation is then applied separately to each of the fields $\meanBB^{pq}$,
where $p=1$ or 2 and $q=c$ or $s$ label different test fields.
Brandenburg et al.\ (2008a,b) use the four test fields
\EQ
\meanBB^{1ck}=B_0(\cos kz, 0, 0), \quad \meanBB^{1sk}=B_0(\sin kz, 0, 0),
\EN
\EQ
\meanBB^{2ck}=B_0(0, \cos kz, 0), \quad \meanBB^{2sk}=B_0(0, \sin kz, 0),
\EN
where the third superscript $k$ has been added to denote the wavenumber,
and $B_0$ is a normalization factor.
The response to each test field, $\bb^{pqk}$, is found by solving
\Eq{Testfield_b1}.
In this way, one finds $\meanEMF^{pqk}=\overline{\uu\times\bb^{pqk}}$
and obtains $4\times2$ equations,
\EQ
\meanemf_i^{pqk}=\alpha_{ij}\meanB_j^{pqk}-\eta_{ij}\mu_0\meanJ_j^{pqk},
\label{ansatz2}
\EN
for the $4+4$ unknowns, $\alpha_{ij}$ and $\eta_{ij}$, for $i=1,2$
and $j=1,2$.
These eight unknowns are obtained as
\EQ
\pmatrix{\alpha_{ij}\cr\eta_{ij3}k}=B_0^{-1}
\pmatrix{\cos kz & \sin kz\cr-\sin kz & ~~\cos kz}
\pmatrix{\meanemf_i^{jck}\cr\meanemf_i^{jsk}},
\label{mean_est_E_standard}
\EN
where the rank-3 tensor $\eta_{ij3}$ is related to the rank-2 tensor
in \Eq{ansatz2} via $\eta_{ij}=\eta_{ik3}\epsilon_{jk3}$.
Note that the result is independent of the value of $B_0$.
For stationary isotropic homogeneous turbulence we have
constant values of $\alpha_{11}=\alpha_{22}\equiv\alpha$ and
$\eta_{11}=\eta_{22}\equiv\etat$, except for statistical fluctuations
resulting from finite computational volumes.

The test-field method for a passive scalar works analogously
(Brandenburg et al.\ 2009).
The concentration per unit volume $C$ obeys the equation
\EQ
{\partial C\over\partial t}=-\nab\cdot(\UU C)+\kappa\nabla^2C,
\label{dCdt}
\EN
and the evolution of the mean concentration $\meanC$ is obtained by
averaging \Eq{dCdt}, which yields
\EQ
{\partial\meanC\over\partial t}=-\nab\cdot(\meanUU\,\meanC+\meanFFFF)
+\kappa\nabla^2\meanC.
\label{dmeanCdt}
\EN
The test scalar equation is obtained by subtracting \Eq{dmeanCdt} from
\Eq{dCdt}, which yields
\EQ
{\partial c\over\partial t}=
-\nab\cdot(\meanUU c+\uu\meanC+\uu c-\overline{\uu c})
+\kappa\nabla^2c.
\label{dcdt}
\EN
In order to obtain $\kappat$, one uses the test scalars
\EQ
\meanC^{ck}=C_0\cos kz,\quad\meanC^{sk}=C_0\sin kz,
\EN
where $q=c$ or $s$ denotes the spatial dependence of the test scalar and,
again, an additional superscript $k$ denotes the wavenumber,
while $C_0$ is a normalization factor.
For each test scalar, we obtain a separate evolution equation for $c^{qk}$.
In this way, we calculate the fluxes,
$\meanFFFF^{qk}=\overline{\uu c^{qk}}$, and compute
the three components of $\kappa_{i3}$:
\EQ
\kappa_{i3}=-\left\langle-\sin kz\meanFFF_i^{ck}+\cos kz\meanFFF_i^{sk}\right\rangle_{z}/kC_0,
\label{Fz}
\EN
for $i=1, ..., 3$, where $\bra{\,}_z$ denotes a $z$ average.
Again, the values of $\kappa_{i3}$ are independent of the normalization
constant $C_0$.
For stationary isotropic homogeneous turbulence we have, except for
statistical fluctuations, constant $\kappa_{ij}=\kappat\delta_{ij}$.

By applying the test-field and test-scalar methods to a range of
different wavenumbers $k$, it was possible to assemble two full integral
kernels in space (Brandenburg et al.\ 2008b, 2009) and hence to take the
effects of finite scale separation into account.
In the following, we proceed analogously by applying the test-field and
test-scalar methods to a range of different frequencies to assemble
two full integral kernels in time and hence to take memory effects into
account.

\subsection{Determination of the kernels}

As is common in linear response theory, all integral kernels vanish for $t<0$.
Therefore the integrations in \Eqs{convolutiont1}{meanEMF}
extend effectively only to $t'=t$.
In order to determine these kernels numerically, we can either calculate
them directly by imposing $\delta$ function-like variations of the test fields,
or we can use the fact that a convolution corresponds to a multiplication
in spectral space, i.e.\
\EQ
\tildeFFFF(\omega)=-\tilde\kappat(\omega)\tildeGG(\omega),
\EN
where
\EQ
\tilde\kappat(\omega)=\int\dd t\, e^{\ii\omega t}\hat\kappat(t)
\label{kappaFT}
\EN
is the Fourier transform  of $\hat\kappat(t)$.

A multiplicative relation between $\meanFFFF$ and $\meanGG$ applies
also to the Laplace transform of these functions with
\EQ
\tildeFFFF(s)=-\tilde\kappat(s)\tildeGG(s),
\EN
where 
\EQ
\tilde\kappat(s)=\int_0^\infty\dd t\, e^{-st}\hat\kappat(t)
\EN
is now the Laplace transformation of $\hat\kappat(t)$.

We introduce an additional superscript $\omega$ for the cases where the
test fields or concentrations have $\cos\omega t$ time dependence
and superscript $s$ for the cases where the test fields or
concentrations have $\exp st$ time dependence.
The superscripts or the explicit time dependence are sometimes suppressed.
In most of the cases we use test fields with a sinusoidal spatial
dependence with wavenumber $k=k_1$.
However, it is sometimes useful to vary also the value of $k$.
In these cases, we also add the superscript $k$.

The multiplicative relations above imply
that for an oscillatory perturbation with a single
frequency there is a multiplicative relation between $\meanGG^{qk\omega}(t)$
and $\meanFFFF^{qk\omega}(t)$, where the first superscript denotes the frequency;
see Appendix~\ref{convolutiont}.
In general, $\kappat$ is a tensor, but in the following we restrict
ourselves to determining only one of its components, namely the one
relating the $z$ components of $\meanFFFF(t)$ and $\meanGG(t)$ to
each other.
We therefore assume $\meanGG(z,t)=(0,0,\meanG)$, where
$\meanG=\partial C/\partial z$.
The different test scalars $C^{qk\omega}$ are denoted by superscripts
$c$ and $s$ for spatial dependences proportional to $\cos kz$ and $\sin kz$,
so we have
\EQ
\meanC^{ck\omega}=C_0\cos kz\cos\omega t,\quad
\meanC^{sk\omega}=C_0\sin kz\cos\omega t,
\EN
for oscillatory test fields, and
\EQ
\meanC^{cks}=C_0\cos kz\exp st,\quad
\meanC^{sks}=C_0\sin kz\exp st,
\EN
for exponentially growing or decaying test fields.
For each value of $\omega$ we determine the resulting $z$ component of the
flux, $\meanFFF_\omega(t)$.
As shown in \Eq{appAkt} of \App{convolutiont}, we can
calculate the response kernel as
\EQ
\tilde\kappat(k,\omega)=-2G_0^{-1}\left\langle e^{\ii\omega t}
\meanFFF^{k\omega}(t)\right\rangle_t\,,
\label{mean_eiot_F}
\EN
where the subscript $t$ behind an angular bracket denotes a time average.
Note that $\tilde\kappat(k,\omega)$ is complex such that its real part
is symmetric about $\omega=0$, while the imaginary part is antisymmetric.
In other words, it obeys the Kramers relation,
$\tilde\kappat(k,-\omega)=\tilde\kappat^*(k,\omega)$,
where the asterisk denotes complex conjugation; see, e.g.,
Moffatt (1978) and Krause \& R\"adler (1980).
In our case, in addition, $\tilde\kappat$ is symmetric in $k$.

Analogous relations apply to $\tildeFFFF(s)$ and $\tildeGG(s)$.
In this case, \Eq{mean_eiot_F} is modified to
\EQ
\tilde\kappat(k,s)=-G_0^{-1}\left\langle e^{-st}
\meanFFF^{ks}(t)\right\rangle_t\,.
\label{mean_est_F}
\EN

As discussed in \Sec{Standard}, our test fields allow us to pick out
each tensor component of $\alpha_{ij}$ and $\eta_{ij}$ separately.
We therefore define time-dependent test fields
\EQ
\meanBB^{pqk\omega}=\meanBB^{pqk}\cos\omega t,\quad\mbox{and}\quad
\meanBB^{pqks}=\meanBB^{pqk}\exp st,
\EN
where the time-independent test fields $\meanBB^{pqk}$ were defined in
\Eq{ansatz2}.
Owing to variations of the form $\sin  kz$ and $\cos kz$ one multiplies
with the inverse of a rotation matrix,
\EQ
\pmatrix{\tilde\alpha_{ij}(k,s)\cr\tilde\eta_{ij3}(k,s)k}=\left\langle e^{-st}
\pmatrix{\cos kz & \sin kz\cr-\sin kz & ~~\cos kz}
\pmatrix{\meanemf_i^{1jks}\cr\meanemf_i^{2jks}}\right\rangle_t,
\label{mean_est_E}
\EN
where the matrix above results from the choice of the sinusoidal test
fields; see Sur et al.\ (2008) for details.
An analogous equation applies also to the case of oscillatory test fields
where $s$ is replaced by $-\ii\omega$, so we write
\EQ
\pmatrix{\tilde\alpha_{ij}(k,\omega)\cr\tilde\eta_{ij3}(k,\omega)k}=\left\langle e^{\ii\omega t}
\pmatrix{\cos kz & \sin kz\cr-\sin kz & ~~\cos kz}
\pmatrix{\meanemf_i^{1jk\omega}\cr\meanemf_i^{2jk\omega}}\right\rangle_t,
\label{mean_eot_E}
\EN
keeping in mind that a tilde has been used to indicate both Fourier and
Laplace transformation.

\section{Preliminary considerations}
\label{Preliminary}

Before entering the numerical determination of the integral kernels
let us consider a current approach that captures memory effects,
as well as its simplest extension.
This will later serve us with a useful fit formula for the more
complicated cases.

\subsection{Expectations from the $\tau$ approximation}
\label{TauApprox}

We use the term $\tau$ approximation here in the form introduced by
Blackman \& Field (2002, 2003, 2004).
The essence of the $\tau$ approximation is to write down
evolution equations for second order correlations
such as $\overline{\uu c}$ and $\overline{\uu\times\bb}$.
This results in triple correlations that are not
omitted, as in the first-order smoothing approximation
(FOSA), but are instead approximated by a closure hypothesis.
In the $\tau$ approximation, one replaces the triple correlations
by quadratic correlations divided by a relaxation time $\tau$
(Vainshtein \& Kitchatinov 1983; Kleeorin et al.\ 1996).
This timescale is expected to be comparable to the turnover time
of the turbulence.

Blackman \& Field (2002, 2003, 2004) were the first to {\it retain} the
time derivative in the evolution equations for $\overline{\uu c}$ and
$\overline{\uu\times\bb}$.
This means that the Fickian diffusion approximation of \Eq{FickianFE}, i.e.\
$\meanFFFF=-\tilde\kappatz\meanGG$, with $\tilde\kappatz=\tau\overline{u_z^2}$
for one-dimensional diffusion in the $z$ direction, is generalized to
\EQ
\left(1+\tau{\partial\over\partial t}\right)\meanFFFF=-\tilde\kappatz\meanGG.
\label{tauapprox}
\EN
This implies that the Fourier-transformed integral kernel is
\EQ
\tilde\kappat(\omega)={\tilde\kappatz\over1-\ii\omega\tau}
={\tau\overline{u_z^2}\over1-\ii\omega\tau}.
\label{kappa_omega}
\EN
(Any $k$ dependence is here ignored.)
In real space, this expression for $\tilde\kappat(\omega)$ corresponds
to the integral kernel
\EQ
\hat\kappat(t)=\int_{-\infty}^\infty{\dd\omega\over2\pi}
e^{-\ii\omega t}{\tau\overline{u_z^2}\over1-\ii\omega\tau}
=\overline{u_z^2}\,\Theta(t)\,e^{-t/\tau},
\EN
where the integral has been solved as a contour integral around the
pole at $\omega=-\ii/\tau$, and $\Theta$ is the Heaviside step function
with $\Theta(t)=1$ for $t>0$ and 0 otherwise.
In the limit $\tau\to0$, the exponential
function reduces to $\tau\delta(t)$, so
\EQ
\hat\kappat(t)\to\tau\overline{u_z^2}\,\delta(t)
=\tilde\kappatz\delta(t)\quad\mbox{(for $\tau\to0$)},
\EN
and one recovers the usual prediction in which turbulent diffusion can
be treated as a multiplicative enhanced diffusion coefficient.

Similar considerations also apply to the case with magnetic fields, where
$\meanEMF$ is essentially being replaced by $(1+\tau\partial_t)\meanEMF$.
For exponentially growing solutions, one would therefore expect that the
actual growth rate $\lambda$ is reduced by a factor $(1+\lambda\tau)^{-1}$.
However, this expectation may be too naive and will need to be reconsidered
in this work.

\begin{figure*}[t!]
\centering\includegraphics[width=\textwidth]{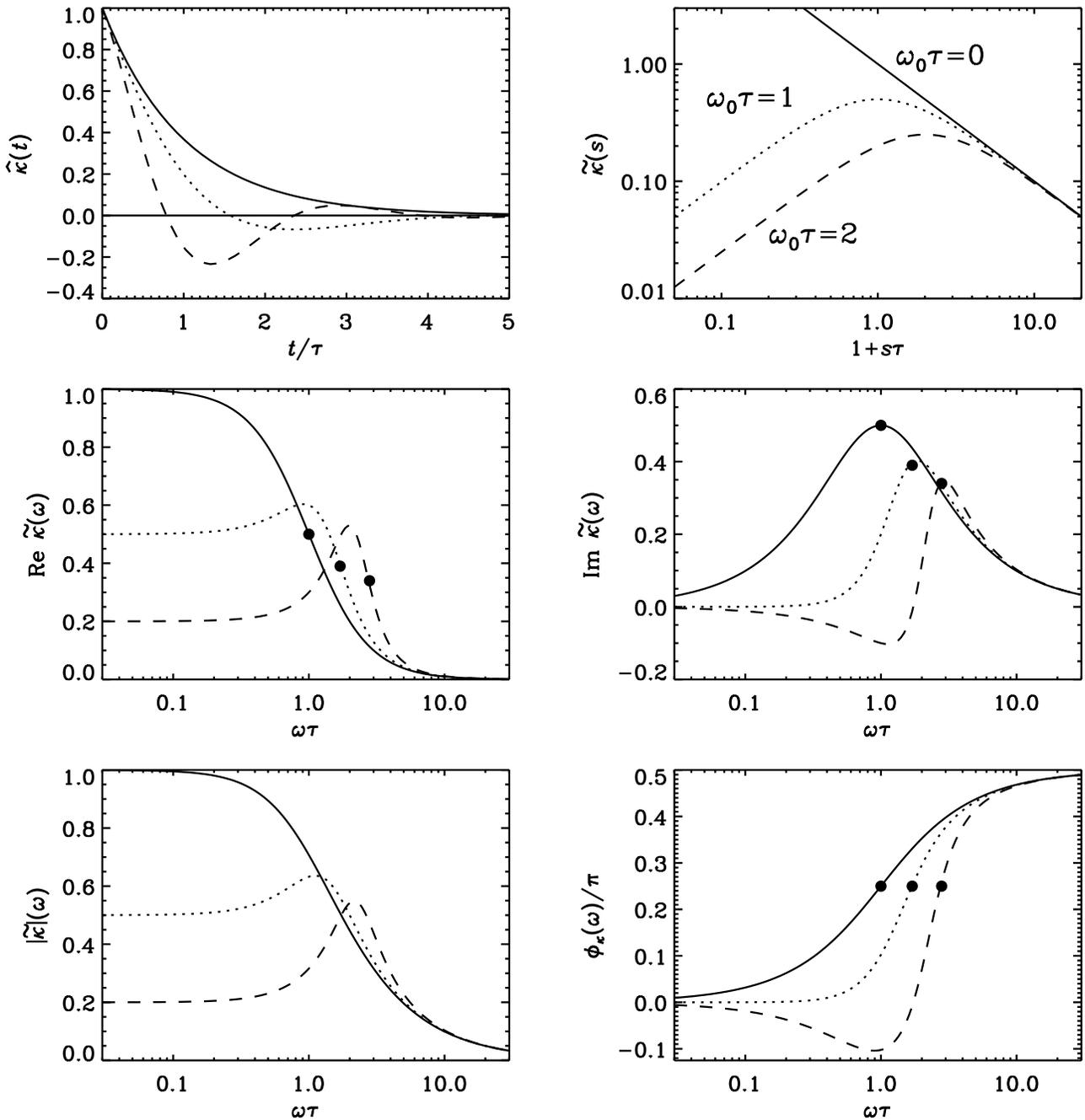}\caption{
Plots of the model integral kernel given by \Eq{predic} for 
$\omega_0\tau=0$ (solid lines), 1 (dotted lines), and 2 (dashed lines),
compared with its Laplace transform ($\tilde\kappat(s)$),
the real and imaginary parts of $\tilde\kappat(\omega)$,
its modulus $|\tilde\kappat|$ and phase $\phi_\kappa$.
The positions where ${\rm Re}\,\kappa_{\rm t}={\rm Im}\,\kappa_{\rm t}$
are marked with filled symbols in three relevant panels.
}\label{o0_dependence}\end{figure*}

A useful diagnostic for the applicability of \Eq{kappa_omega} is that the value of
$\omega$ where ${\rm Re}\,\kappa_{\rm t}={\rm Im}\,\kappa_{\rm t}$ is also the value of $\omega$
where $\dd\,{\rm Im}\,\kappa_{\rm t}/\dd\omega =0$
(i.e.\ where the phase is $\pi/4$, see \Fig{o0_dependence}, final panel).
It will turn out that this property is not always obeyed.

\subsection{Effects beyond FOSA and $\tau$ approximation}
\label{EffectsAdvection}

While a $\delta$ function perturbation is disadvantageous numerically, it can be illuminating.
If we impose on a flow with $\meanUU=\bf{0}$
a test-field $\meanC$ with a $\delta(t)$ time dependence, then
the value of $c(0)$ depends only on the $\nab \cdot (\uu \meanC)$ term
in \Eq{dcdt}.
For $t>0$, \Eq{dcdt} reduces then to:
\EQ
{\partial c\over\partial t}=
-\nab\cdot(\uu c-\overline{\uu c})
+\kappa\nabla^2c\quad\mbox{($t>0$)}.
\label{deltafc}
\EN
Such a $\delta$ perturbation then launches fluctuating fields 
which evolve according to an equation
similar to \Eq{deltafc}.  In passive scalar or kinematic dynamo cases, 
the evolution of the fluctuating field depends only on $\uu$, which is independent of the  
fluctuating field.  The fluctuating fields will decay exponentially according to turbulent or
micro-physical diffusion or resistivity, but they will generate
$\meanFFFF$ (or $\meanEMF$ in the magnetic case) for as long as they survive.
It is the finite lifetime of the fluctuating fields that is at the
physical core of this memory effect.

In the passive scalar case we consider $\meanFFFF=\overline{\uu c}$, to which
the term $\nab \cdot (\overline{\uu c})$ in \Eq{deltafc} does not 
contribute because $\overline {\uu\nab\cdot (\overline{\uu c})}=0$.
If the spatial dependence of our 
test scalar $\meanC$ is sinusoidal and 
lies only along a direction $\hat z$, then $c(0)$ will also have only sinusoidal
behavior in that direction.
If we imagine that the initial $c(0)$ is proportional to $\sin kz$, then,
in the absence of other effects, two counter-propagating vertical streams 
with $u_z=\pm u$ (assuming $\meanUU=0$) will generate an advective 
sinusoidal signal from the $\nabla_z(u_z c)$ term of \Eq{deltafc}:
 \EQ
\overline{u_z\nabla_z(u_zc)}(t,z)=2u^2k \cos kz\cos \omega_0t,
 \EN
where $\omega_0=ku$.
In a turbulent system the above analysis can only be done for times shorter
than or comparable to a turbulent correlation time $\tau \sim 1/ku$.  
For times larger than a turbulent correlation time,
the standard $e^{-t/\tau}$ diffusion term will be important.  A
``turbulent" diffusion is formally possible even in steady flows,
but it will be just the microscopic diffusion.  

We can combine the short timescale
advective (oscillatory) and longer timescale diffusive (exponential) effects 
by a simple multiplication:
 we expect
the form for $\hat{\kappat}(t)$ to be similar to
 \EQ
 \hat{\kappat}(t) \simeq u^2\,\Theta(t)\,e^{-t/\tau}\cos\omega_0 t.
 \label{predic}
 \EN
 Note that in a turbulent system we expect $\omega_0 \sim 1/\tau$ on dimensional
 grounds and so the above analysis is not rigorous. 
 However, as we will see in \Sec{sectionresults} this form fits the results reasonably well.
 In the Fourier space this becomes
\EQ
{\tilde\kappat(\omega)\over\tilde\kappa_{t0}}
={1-\ii\omega\tau\over(1-\ii\omega\tau)^2+\omega_0^2\tau^2},
\label{response_omega_osc}
\EN
where $\tilde\kappa_{t0}=\tau u^2$ has been assumed, although this
prefactor may not be accurate for $\omega_0\neq0$.
The corresponding Laplace transform is
\EQ
{\tilde\kappat(s)\over\tilde\kappa_{t0}}
={1+s\tau\over(1+s\tau)^2+\omega_0^2\tau^2}.
\label{response_ss_osc}
\EN
In the limit $\omega_0\to0$ these expressions coincide with those
of \Sec{TauApprox}.
In \Fig{o0_dependence} we plot various representations of the integral kernel
for different values of $\omega_0\tau$.

In order to assess whether the proposed extension to capturing memory
effects is viable, we shall use \Eqs{response_omega_osc}{response_ss_osc}
as fit formulae to determine the value of $\omega_0$ and to find out how
it depends on other aspects of the model such as the P\'eclet number
and wavenumber of the mean concentration.

In the absence of a detailed analogous motivation for $\alpha$ and $\etat$
we shall use in this paper \Eqs{response_omega_osc}{response_ss_osc}
as fit formulae also in the magnetic case.
In this case, we use these formulae for $\alpha$ and $\etat$ and
add corresponding subscripts $\alpha$ and $\eta$ to $\tau$ and
$\omega_0$, where it replaces the subscript 0, i.e.\ we write
\EQ
{\tilde\alpha(\omega)\over\tilde\alpha_{0}}
=A_{\alpha}{1-\ii\omega\tau_\alpha\over(1-\ii\omega\tau_\alpha)^2
+\omega_\alpha^2\tau_\alpha^2},
\label{alpha_omega_osc}
\EN
\EQ
{\tilde\etat(\omega)\over\tilde\eta_{t0}}
=A_{\eta}{1-\ii\omega\tau_\eta\over(1-\ii\omega\tau_\eta)^2
+\omega_\eta^2\tau_\eta^2}.
\label{etat_omega_osc}
\EN
Again, we expect $\omega_\alpha\tau_\alpha$ and $\omega_\eta\tau_\eta$
to be of order unity, but in this paper we allow them to be
adjustable parameters.  
Further, we use $A_{\alpha}$ and $A_{\eta}$ as further
fit parameters, modifying the amplitude.
The relaxation times $\tau$ and $\omega_0^{-1}$ and values derived from them
such as $\tilde\alpha_{t0}$ are
merely characteristic times, and we do not attempt to laboriously average
over the true values.

Note that the above form for the kernel, \Eq{predic},
is the simplest extension of the 
$\tau$ approximation that qualitatively fits our simulation results.
From that perspective, we replace \Eq{tauapprox} by:
\EQ
\left(1+\omega_0^2\tau^2+2\tau{\partial\over\partial t}
+\tau^2{\partial^2 \over \partial t^2} \right)\meanFFFF
=-\tilde\kappat \left(1+\tau{\partial \over \partial t} \right)\meanGG.
\label{tauapprox2}
\EN
Note also that, unlike \Eq{kappa_omega}, for \Eq{response_omega_osc},
the value of $\omega$ where the slope of the imaginary component is zero is not the
same as the value of $\omega$ where the phase is $\pi/4$ (see the end of \Sec{TauApprox}).

\begin{figure}[t!]
\centering\includegraphics[width=\columnwidth]{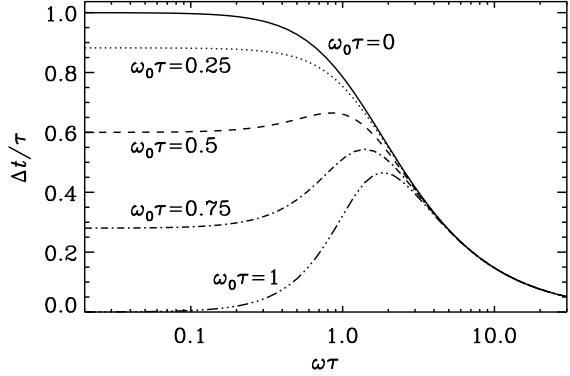}\caption{
Normalized time lag $\Delta t/\tau$ versus $\omega\tau$ for
different values of $\omega_0\tau$.
Note the development of a peak near $\omega\tau=1$ as $\omega_0\tau$
is increased.
}\label{phase}\end{figure}

As shown in \Eq{TimeDelay} of \App{convolutiont}, for monochromatic
mean fields a phase shift $\phi_\kappa$ leads to a time lag
\EQ
\Delta t=\phi_\kappa(\omega)/\omega,
\EN
so the flux $\meanFFF_\omega(t)$ depends only on the
mean concentration gradient at time $t-\Delta t$ and is
given by $-|\tilde\kappa|\,\meanG(t-\Delta t)$.
For the response function given by \Eq{response_omega_osc}, the time lag is
\EQ
{\Delta t\over\tau}
={\phi_\kappa(\omega)\over\omega\tau}
={1\over\omega\tau}{\rm \arctan}\left[\omega\tau
{1+(\omega^2-\omega_0^2)\tau^2\over1+(\omega^2+\omega_0^2)\tau^2}\right],
\EN
which always vanishes for large values of $\omega$
and can have a peak near $\omega\tau=1$ for $\omega_0>\omega_0^*$
with $\omega_0^*\tau\approx0.3273$; see \Fig{phase}.

\section{Simulations}
\label{sim}

We consider two types of flows.
For test purposes and comparison with earlier work
described in \Sec{Background}
we use the Roberts flow.
The Roberts flow is
given by
\EQ
\UU=\kf\psi\zzz-\zzz\times\nab\psi,
\label{Rob1a}
\EN
with
\EQ
\psi=(u_0/k_0)\cos k_0x\cos k_0y
\label{Rob2a}
\EN
and $k_{\rm f} \equiv \sqrt{2}k_0$ so that $k_{\rm{f}}^2=k_x^2+k_y^2$,
where $k_x=k_y=k_0$ is the wavenumber of the flow in the $xy$ plane.
This flow is capable of dynamo action once the magnetic Reynolds number,
\EQ
\Rm=\urms/\eta k_0,
\label{Rmdef}
\EN
exceeds a critical value, $\Rm\ge\Rmc\equiv5.52$.
[We recall that our test fields have spatial dependence given
by $k_1$, i.e.\ the smallest wavenumber that fits in the box.
We note further that for $\Rm\leq70$ the most unstable wavenumber
of the field that fits into the box is still $|k_z|=k_1$, where $k_z$
was defined in \Eq{disper} and this agrees with the wavenumber
of the test fields.
Note however that, for $\Rm=100$, for example, the most unstable mode
would have $|k_z|=2k_1$.]

The other alternative is forced turbulence.
In that case we consider an isothermal equation of state with constant speed
of sound, $c_{\rm s}$, and solve the momentum and continuity equations
\EQ
{\partial\UU\over\partial t}=-\UU\cdot\nab\UU-c_{\rm s}^2\nab\ln\rho
+\ff+\rho^{-1}\nab\cdot2\rho\nu\SSSS,
\label{momentum}
\EN
\EQ
{\partial\rho\over\partial t}=-\nab\cdot(\UU\rho),
\label{continuity}
\EN
where $\ff$ is a random forcing function consisting of circularly
polarized plane waves with positive helicity and random direction and phase,
$\SSSS$ is the traceless rate-of-strain tensor.
The length of the wavevector of the forcing function,
$|\kk_{\rm f}|$, is chosen to be in a narrow band around an average
wavenumber $k_{\rm f}$.
We adjust the strength of the forcing such that the flow remains
clearly subsonic (mean Mach number is around 0.1).
The details of the forcing function used in the present work
can be found in Appendix A of Brandenburg \& Subramanian (2005).
For forced turbulence we define $\Rm=\urms/\eta\kf$.

We consider a domain of size $L_x\times L_y\times L_z$.
In all cases, we take $L_x=L_y=L_z=2\pi/k_1$.
The ratio $k_{\rm f}/k_1$ is referred to as the scale separation ratio.
Our model is characterized by the choice of fluid and magnetic Reynolds
numbers as well as the P\'eclet number, based here on the wavenumber $\kf$.
The magnetic Reynolds number was defined in \Eq{Rmdef}.
The fluid Reynolds and P\'eclet numbers are defined analogously,
\EQ
\Rey=\urms/\nu\kf,\quad
\Pe=\urms/\kappa\kf,
\EN
where the magnetic diffusivity $\eta$ is replaced by the viscosity $\nu$
and the molecular diffusivity $\kappa$, respectively.

We present the results in non-dimensional form by normalizing
$\tilde{\kappat}(\omega)$, analogously to earlier work
(Brandenburg et al.\ 2008a), by
\EQ
\tilde\kappatz=\tau\overline{u_z^2}=\half\tau\urms^2
\quad\mbox{(for the Roberts flow)}.
\label{kappatz}
\EN
For turbulent flows, $\tau$ is proportional to the turnover time,
$(\urms\kf)^{-1}$.
However, in the limit of low P\'eclet number, microscopic diffusion
becomes important and dominates over the triple correlation terms.
This means that the effective $\tau$ is given by the microscopic
diffusion time $\tau=(\kappa\kf^2)^{-1}$.

We define the Strouhal number as $\mbox{St}=\tau u_{\rm rms}k_{\rm f}$
and can then write $\tau$ as
\EQ
\tau=\St/(u_{\rm rms}k_{\rm f}).
\label{St}
\EN
The value of $\St$ characterizes the flow field.
For turbulent flows of the form discussed in the present paper
its value is of order unity (Brandenburg et al.\ 2004).
Later in this paper we shall allow $\St$ to be a fit parameter.
We present the results for $\alpha$ and $\eta_{\rm t}$
by normalizing, depending on the nature of the flow with
\EQ
\tilde\alpha_0=-\half u_{\rm rms},\quad
\tilde\eta_{\rm t0}=\half u_{\rm rms}k_{\rm f}^{-1}
\quad\mbox{(Roberts flow)}
\EN
and
\EQ
\tilde\alpha_0=-\onethird u_{\rm rms},\quad
\tilde\eta_{\rm t0}=\onethird u_{\rm rms}k_{\rm f}^{-1}
\quad\mbox{(3D turbulence)}.
\EN
as was done in Brandenburg et al.\ (2008a).

\section{Results}
\label{sectionresults}

Our choice of \Eq{FickianFE} results in transport coefficients
that depend on the wavenumber of the mean fields.
 Throughout this section we will assume that our mean fields
 vary spatially according to $k_z=k_1$ unless otherwise specified.
For simplicity therefore, we drop the fixed argument $k_z$ in
$\tilde\kappa(k_z,\omega)$, $\tilde\alpha(k_z,\omega)$ and
 $\tilde\etat(k_z,\omega)$, and similarly for
 $\tilde\kappa(k_z,s)$, $\tilde\alpha(k_z,s)$, and $\tilde\etat(k_z,s)$.

\subsection{Passive scalar diffusion}
\label{PSDiff}

\begin{figure}[t!]
\centering\includegraphics[width=\columnwidth]{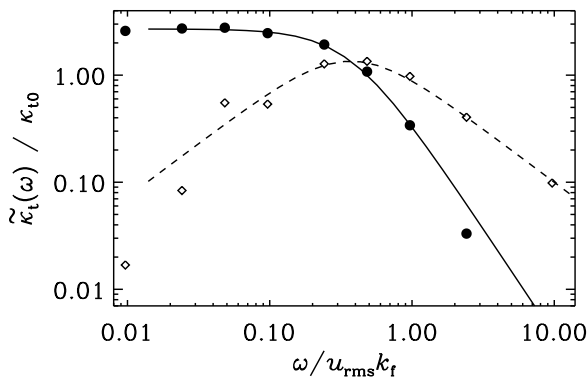}\caption{
Real (solid circles) and imaginary (diamonds) parts of
$\tilde\kappa(\omega)$ for forced turbulence
with $\kf/k_1=2.2$, $\Rey=8$, and $\Pe=40$.  The solid and dashed lines
are a fit using \Eq{kappa_omega} (with 
$\tau$ determined using \Eq{St}, $\St=2.7$ as described in text).
}\label{turb_pscalar_po32}\end{figure}

We now consider solutions of \Eq{dcdt} in the case of a turbulent flow,
and consider first the case with a 
uniform gradient of $\meanC$.
This means that $\meanG^{c0}$ is now constant in space,
with $\meanG^{c0}=G_0(t)$.  
The resulting data agree well with
the expression \Eq{kappa_omega}, where $\tau$ is given by \Eq{St} 
with $\St=2.7$;
see \Fig{turb_pscalar_po32}.
The fact that $\St>1$ should not be too surprising, because such a result
has been obtained earlier for this flow, where $\tau$ was estimated as the
relaxation time in the $\tau$ approximation (Brandenburg et al.\ 2004).

\begin{figure}[t!]
\centering\includegraphics[width=\columnwidth]{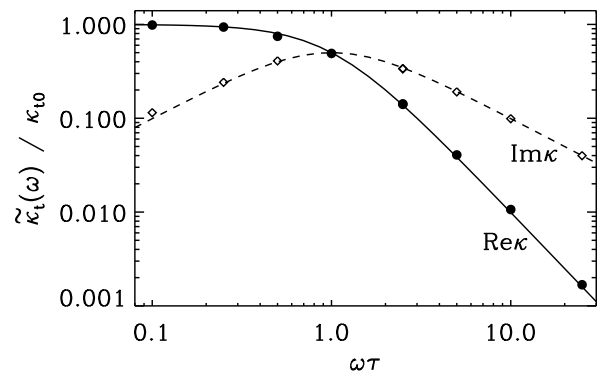}\caption{
Real and imaginary parts 
of $\tilde\kappa(\omega)$ for $\Pe=100$ for the Roberts flow.
Note the perfect agreement with the fit formula \eq{kappa_omega}
using $\tau=1/(\kappa\kf^2)$ and $\tilde\kappatz=\tau\urms^2/2$ (curves).
}\label{kin_pscalar_po16}\end{figure}

The case of the Roberts flow, where $\UU$ is obtained from \Eqs{Rob1a}{Rob2a}
is in some ways more interesting.  In the case of the same uniform gradient
concentration $\meanC^{c0}$,
 the flux can be calculated analytically, as is done in \App{appB}.  As the flow $\UU$
  is steady its correlation timescale is infinite and the only relevant relaxation
  timescale
is the microscopic diffusion time $\tau=(\kappa\kf^2)^{-1}$.
The calculations result in the expression
$\tilde\kappat(\omega)=\kappatz/(1-\ii\omega\tau)$;
see \Eq{kappa_omega}.  This agrees with simulations as
shown in \Fig{kin_pscalar_po16}.

\begin{figure}[t!]
\centering\includegraphics[width=\columnwidth]{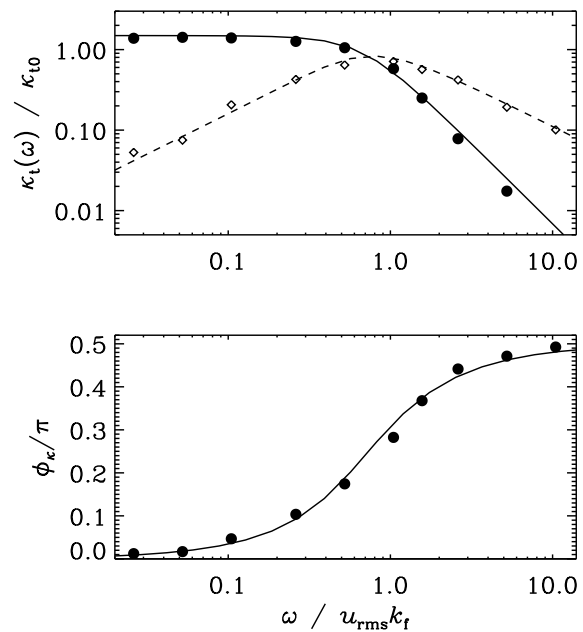}\caption{
Real and imaginary parts of $\tilde\kappa(\omega)$ (upper panel)
and its phase (lower panel) for turbulence at $\Rey=8$ and $\Pe=40$.
The solid and dashed lines are fits.
}\label{turb_pscalar_pto16}\end{figure}

We suggested in \Sec{EffectsAdvection} that advective effects play
a role only when the mean concentration gradient shows a variation
in some direction (i.e., a finite wavenumber), and we should not be surprised
that \Eq{kappa_omega} is adequate to explain the transport of a passive scalar
with zero wavenumber.  The results of \Fig{turb_pscalar_pto16}, where
a turbulent flow is used with $\Rey=8$ and $\Pe=40$, and
a sinusoidal variation of the mean concentration is imposed, are slightly better fitted with
\Eq{response_omega_osc} than with \Eq{kappa_omega}.

The case of the flow $\UU=u_0\zzz\cos x$ with the same sinusoidally
varying concentration is discussed in \App{1D}.
The value of $\omega$ where $\mbox{Re}\,\tilde{\kappa}_t=\mbox{Im}\,\tilde{\kappa}_t$
is not the same as the value of $\omega$ where $\mbox{Im}\,\tilde{\kappa}_t$ has zero slope.
This is implied by \Eq{kappa_omega}, as is discussed in \Sec{TauApprox}.
In \App{1D}, we present a simple one-dimensional model
where the behavior is at odds with \Eq{kappa_omega}, although it can still
be fitted reasonably well with \Eq{response_omega_osc}.
This example also illustrates the difficulty in developing good and simple
fits, as the fit parameters are expected to
depend on the spatial variability (e.g.\ through $\omega_0 \sim ku$).

\subsection{Magnetic fields}

\begin{figure}[t!]
\centering\includegraphics[width=\columnwidth]{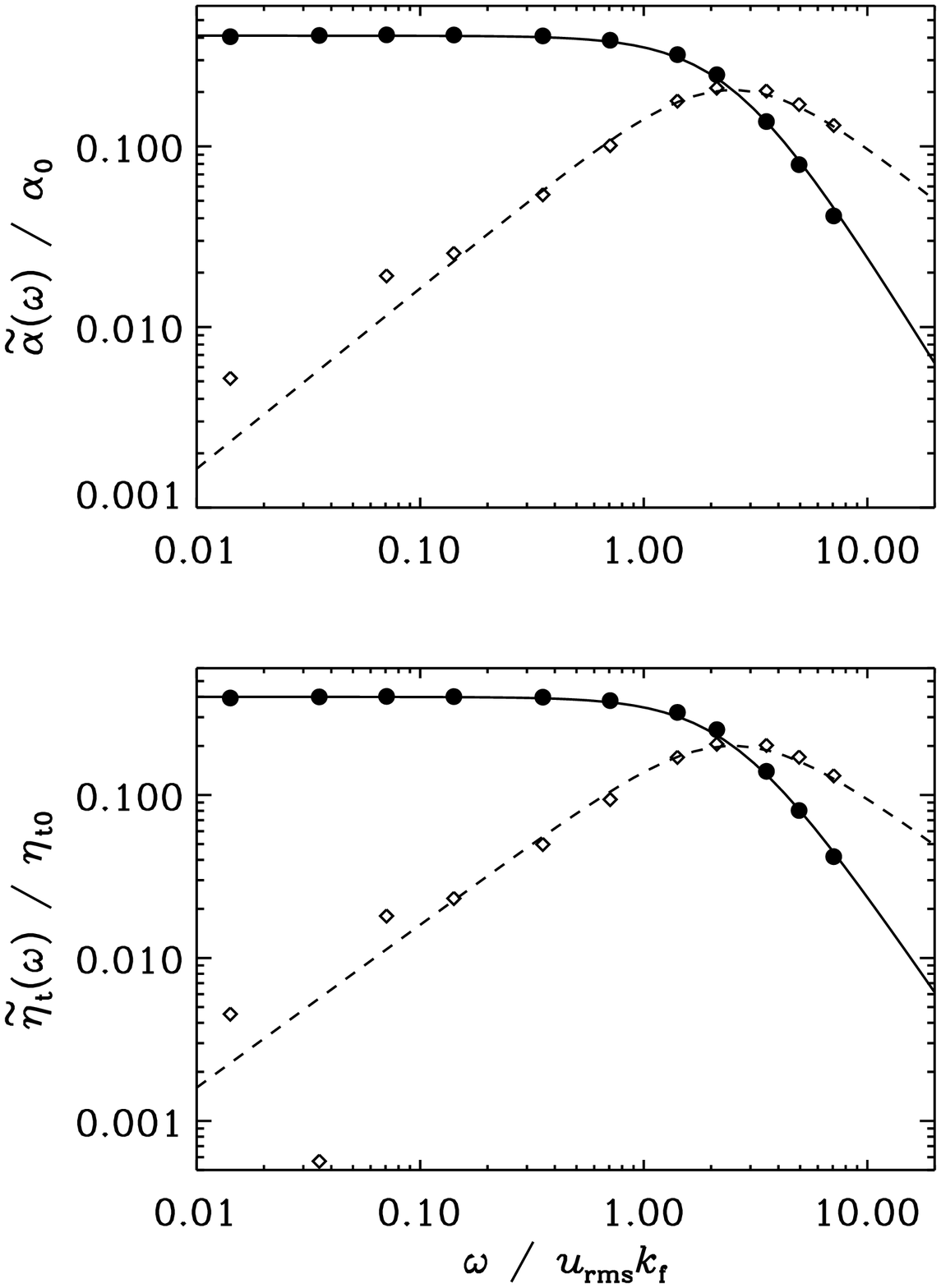}\caption{
Real and imaginary parts of $\tilde\alpha(\omega)$ and $\tilde\etat(\omega)$
for the Roberts flow with $\Rm=1$.
The solid and dashed lines correspond to fits of the form
\Eq{kappa_omega} using \Eq{St} with $\mbox{St}=0.4$.
}\label{scan_Rm1_om_po16}\end{figure}

\begin{figure}[t!]
\centering\includegraphics[width=\columnwidth]{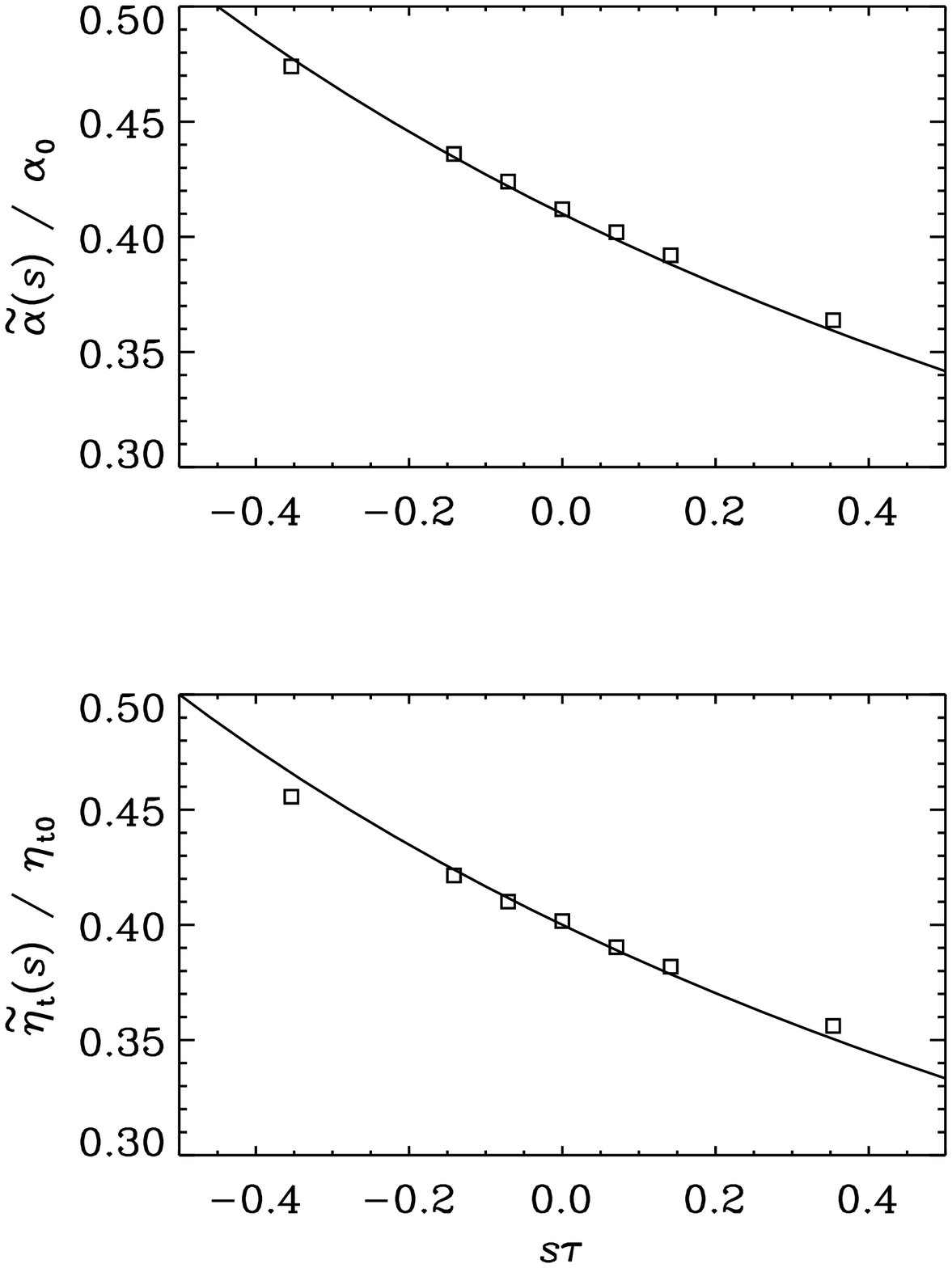}\caption{
Similar to \Fig{scan_Rm1_om_po16}, but for the Laplace-transformed
kernel functions $\tilde\alpha(s)$ and $\tilde\etat(s)$
for the Roberts flow with $\Rm=1$.
The fits are proportional to $\tau/(1+s\tau)$ and correspond to the fits
used in \Fig{scan_Rm1_om_po16}.
}\label{pscan_Rm1_lam}\end{figure}

\begin{figure}[t!]
\centering\includegraphics[width=\columnwidth]{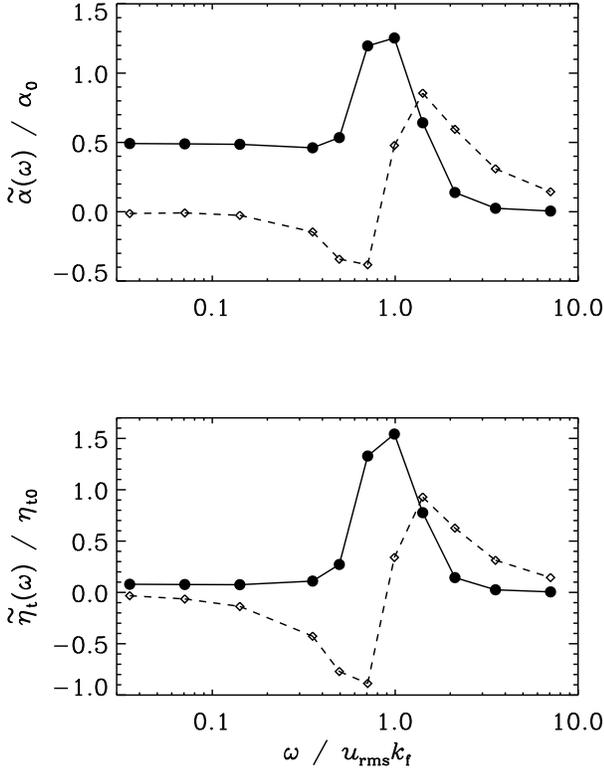}\caption{
Real and imaginary parts of $\tilde\alpha(\omega)$ and $\tilde\etat(\omega)$
for the Roberts flow with $\Rm=10$.  Note that the lines are not analytical fits.
}\label{scan_Rm10_om_po16}\end{figure}

\begin{figure}[t!]
\centering\includegraphics[width=\columnwidth]{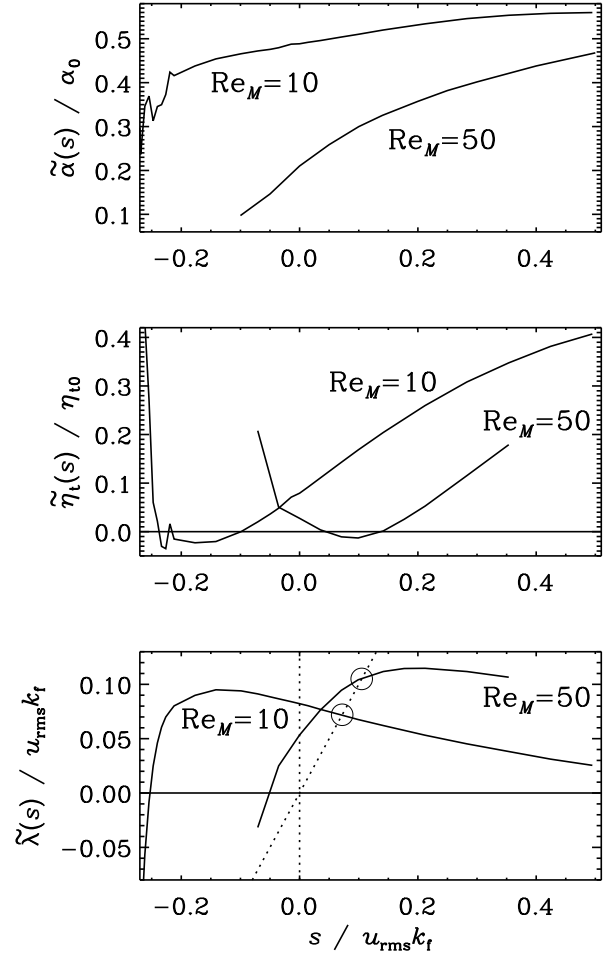}\caption{
Laplace-transformed quantities, 
$\tilde\alpha(s)$, $\tilde\etat(s)$, and $\tilde\lambda(s)$
for the Roberts flow with $\Rm=10$ and $50$.
Note the different signs of the slope at the intersection with the diagonal
(denoted by circles).
}\label{pscan_Rm_both_lam}\end{figure}

For small magnetic Reynolds numbers the functional forms of both
$\tilde\alpha(\omega)$ and $\tilde\etat(\omega)$ are similar to those
in the passive scalar case.
This is demonstrated here for the Roberts flow; see \Fig{scan_Rm1_om_po16},
where $\Rm=1$, which is too small for dynamo action.
\FFig{pscan_Rm1_lam} shows the $s$ dependence for the same case.
However, for $\Rm=10$, which is large enough for dynamo action, the forms of
$\tilde\alpha(\omega)$ and $\tilde\etat(\omega)$ look rather different;
see \Fig{scan_Rm10_om_po16}, which is also for the Roberts flow.
Qualitatively, the data are now closer to \Eq{response_omega_osc},
but a fit would be relatively poor.
Therefore, we cannot rely on a fit to compute the corresponding
Laplace-transformed kernel functions, which are shown in
\Fig{pscan_Rm_both_lam} for $\Rm=10$ and $50$.
Note that, unlike the case of \Fig{pscan_Rm1_lam} for $\Rm=1$,
for $\Rm=10$ and 50, the slope of $\hat\alpha(s)$ is positive.
This is also a feature found by Hori \& Yoshida (2008); see their
Figure~10 for $\Rm=4$, which corresponds to $\Rm=8$ in our definition
of the Roberts flow.

\FFig{pscan_Rm_both_lam} allows us now to assess the error done by applying
the dispersion relation \Eq{disper} with constant values of
$\alpha$ and $\etat$ to cases where $\lambda\neq0$.
A correct procedure would be to use $\tilde\alpha(s)$ and $\tilde\etat(s)$
for $s=\lambda$.
This means that we must calculate
\EQ
\tilde\lambda(s)\equiv\tilde\alpha(s) k_z-\left[\eta+\tilde\etat(s)\right]k_z^2
\EN
for $s=\lambda$.
These points can be obtained from the intersection of $\lambda(s)$
with the diagonal, $\lambda(s)=s$.
For $\Rm=10$ and $50$ these values are at
$\lambda(s=\lambda)\approx0.07\urms\kf$ and $\approx0.11\urms\kf$,
respectively.
By contrast, $\lambda(s=0)\approx0.04\urms\kf$ and $\approx0.07\urms\kf$
for these two values of $\Rm$, respectively.
These values are now in full agreement with those of
$\lambda_{\rm growth}$ seen in \Fig{pscan_Rm}.
This suggests that the reason for the discrepancy between the
two curves in this figure is indeed connected with memory effects.

Let us now turn to the calculation of $\tilde\alpha(\omega)$ and
$\tilde\etat(\omega)$ in the case of turbulence.
In this work we use $\kf/k_1=3$, which is slightly larger than the
values used earlier in the case of a passive scalar.
This value is just large enough to allow for mean field dynamo action
at the minimal wavenumber $k=k_1$ (see Brandenburg et al.\ 2008c).
For $\kf/k_1=2.2$ the scale separation between the scale of the forcing
and that of the domain would be insufficient to allow for large-scale
dynamo action (Haugen et al.\ 2004, Figure 23).

\begin{figure}[t!]
\centering\includegraphics[width=\columnwidth]{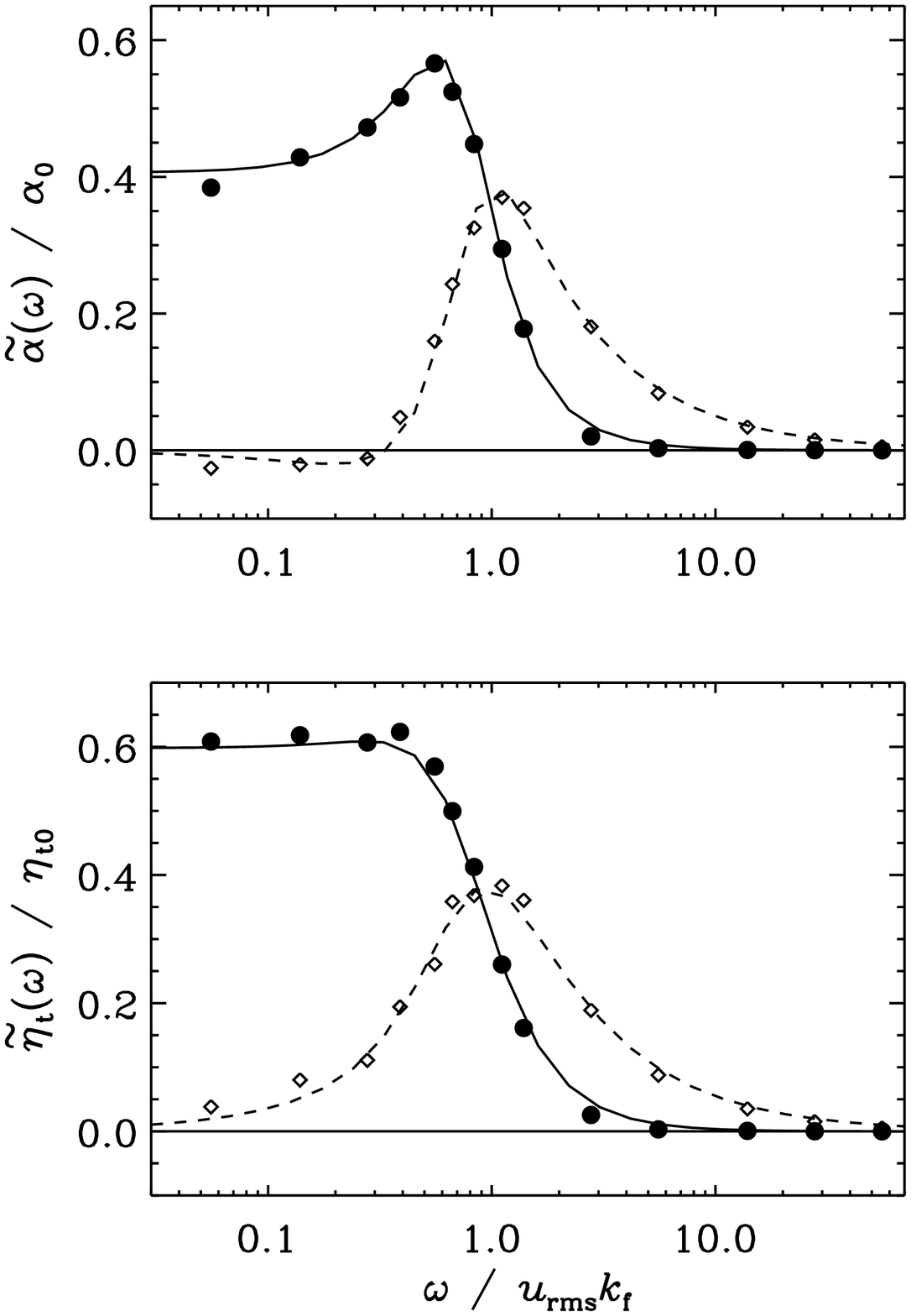}\caption{
Real and imaginary parts of $\tilde\alpha(\omega)$ and $\tilde\etat(\omega)$
for turbulence at $\Rm=22$.
The lines denote fits to \Eqs{alpha_omega_osc}{etat_omega_osc} with
$A_\alpha=1$, $\St_\alpha=2.0$, $\omega_\alpha\tau_\alpha=1.2$, and 
$A_\eta=0.48$, $\St_\eta=1.4$, $\omega_\eta\tau=0.55$, respectively.
}\label{forced_po32}\end{figure}

\begin{figure}[t!]
\centering\includegraphics[width=\columnwidth]{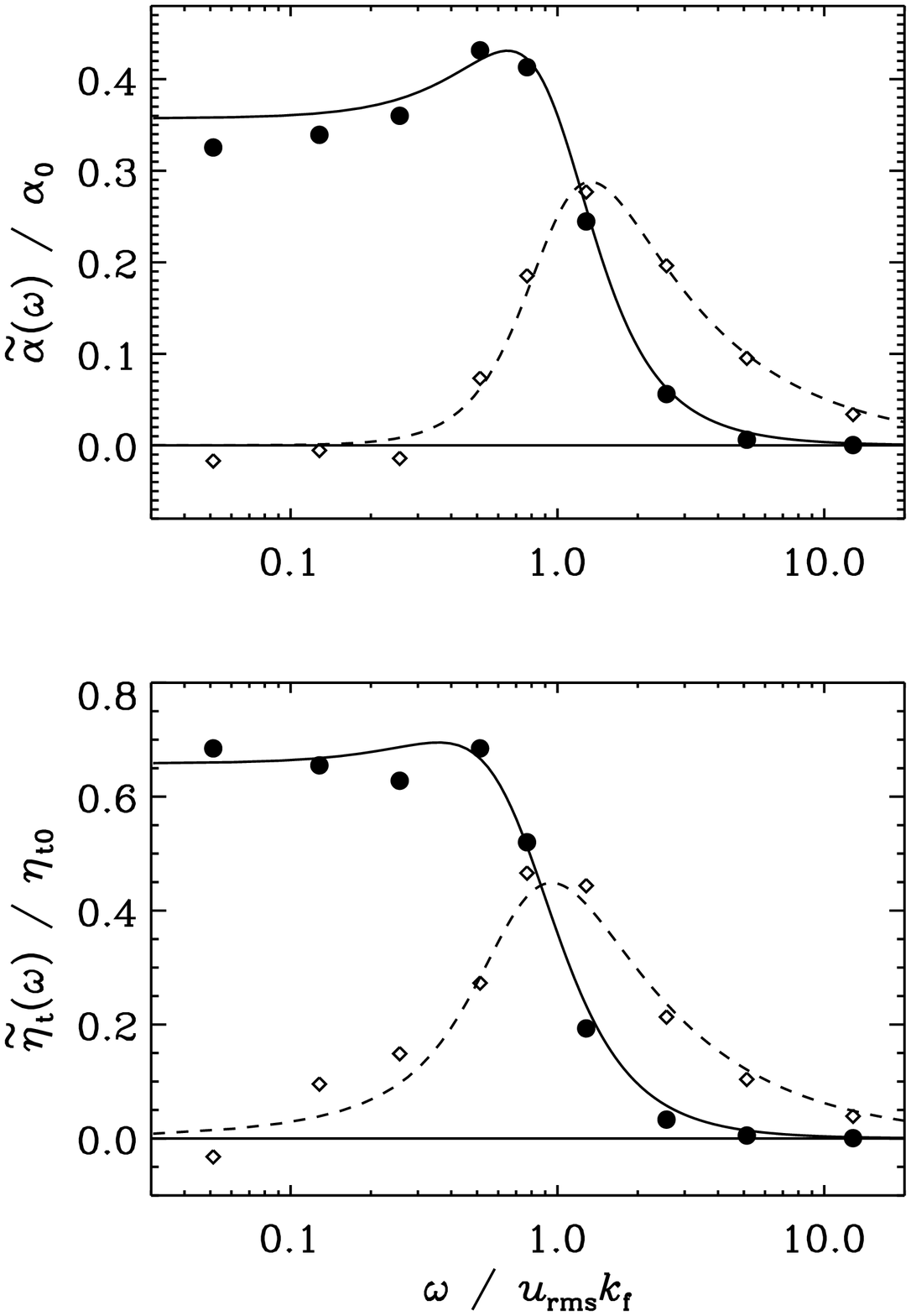}\caption{
Real and imaginary parts of $\tilde\alpha(\omega)$ and $\tilde\etat(\omega)$
for turbulence at $\Rm=90$.
The lines denote fits to \Eqs{alpha_omega_osc}{etat_omega_osc} with
$A_\alpha=1$, $\St_\alpha=1.4$, $\omega_\alpha\tau_\alpha=1$, and 
$A_\eta=1.8$, $\St_\eta=1.7$, $\omega_\eta\tau=0.78$, respectively.
}\label{forced_po128}\end{figure}

\begin{figure}[t!]
\centering\includegraphics[width=\columnwidth]{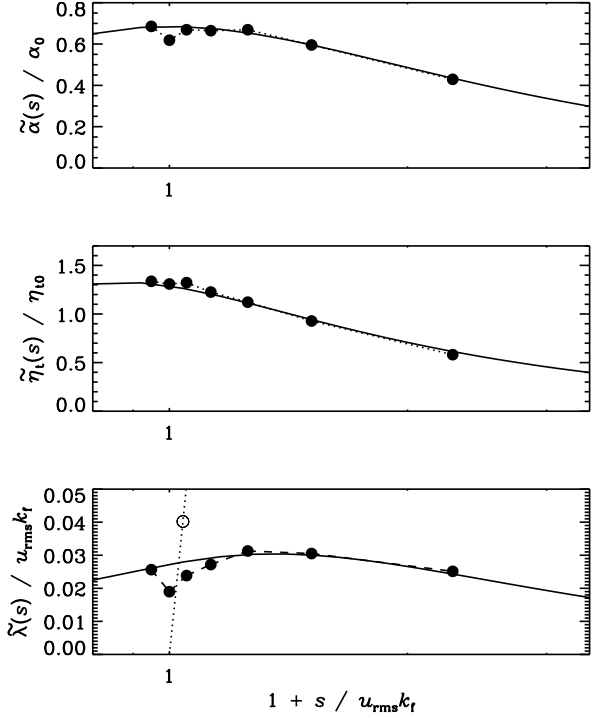}\caption{
Laplace-transformed quantities, 
$\tilde\alpha(s)$, $\tilde\etat(s)$, and $\tilde\lambda(s)$
for forced turbulence at $\Rm=90$.
In the last panel, the diagonal $\lambda=s$ is shown as a dotted line.
The growth rate obtained by solving the three-dimensional induction
equation, which allows for small-scale dynamo action, is indicated by
an open symbol.
}\label{plam128}\end{figure}

\begin{table}[htb]\caption{
Comparison of fit coefficients for $\tilde\alpha(\omega)$ and
$\tilde\etat(\omega)$ for forced turbulence.
}\vspace{12pt}\centerline{\begin{tabular}{lcccccc}
$\Rey$ &
$A_\alpha$ & $\St_\alpha$ & $\omega_\alpha\tau_\alpha$ &
$A_\eta$ & $\St_\eta$ & $\omega_\eta\tau_\eta$ \\
\hline
$22$ & 1.00 & 2.00 & 1.20  & 0.48 & 1.40 & 0.55 \\
$90$ & 1.00 & 1.40 & 1.00  & 1.80 & 1.70 & 0.78
\label{Tsummary}\end{tabular}}\end{table}

By comparing runs of two different magnetic Reynolds numbers, \Fig{forced_po32}
for $\Rm=22$ and \Fig{forced_po128} for $\Rm=90$, we can get some idea
whether the features seen here are artifacts of small values of
$\Rm$, or whether they begin to be of more general significance.
The plots for $\tilde\alpha(\omega)$ and $\tilde\etat(\omega)$
look similar and share the same basic features at both values of $\Rm$,
suggesting that the resulting fits for the response functions
might be robust.
We note that in all cases the phase shows a gradual transition from
0 to $\pi/2$ as $\omega$ increases, but it does not become negative (not shown).
The corresponding fit parameters are summarized in \Tab{Tsummary}.
All the six non-dimensional fit parameters should be of order unity,
and we see that this is indeed the case.
Given that these values have unknown errors connected with the ambiguity
in determining good fits, it is not possible to draw any serious conclusions
from the trends that could be read off the table.

Similar to the case of the Roberts flow,
the fits to the Fourier transformed quantities are not perfect.
Therefore we cannot use the Fourier transform fits to determine
the corresponding Laplace transforms.
In \Fig{plam128} we show the directly determined Laplace transformed
values and compare with the fit inferred from \Fig{forced_po128}.
However, in order to make the fits agree reasonably well, we have
modified the amplitude factors to $A_\alpha=1.37$ and $A_\eta=2.07$.
Note that the agreement is reasonably good, except near $s=0$, where
the actual growth rate is lower than what is inferred from the fit.
This is related to the fact that the actual value of $\tilde\alpha(s)$
near $s=0$ is less than what is predicted by the fit formula.
This suggests that the assumption of similar fit formulae both for
$\alpha$ and $\etat$ may be too simplistic.

As for the Roberts flow, the actual growth rate of the mean-field dynamo
is obtained from the intersection with the diagonal, which is shown as a
dotted line in \Fig{plam128}.
By solving the induction equation for this flow for $\Rm=90$ we find that
the actual growth rate is $0.04\urms\kf$, which is clearly above the point
where $\lambda(s)$ intersects with the diagonal (see the open symbol).
However, this is to be expected, because for $\Rm=90$ there is strong
small-scale dynamo action so the actual growth rate will always exceed
that expected from the mean-field dynamo.
Such a discrepancy was noticed recently in connection with a study of the
dependence of large-scale dynamo action on the magnetic Prandtl number
(Brandenburg 2009).

\section{Discussion}
\label{Discussion}

\subsection{Frequency and growth rate dependence}

An important application of the present results is the determination
of dynamo growth rates.
The usual dispersion relation for isotropic helical turbulence
predicts the growth rate to be
\EQ
\lambda=\alpha k-(\eta+\etat)k^2
\quad\mbox{(for constant $\alpha$, $\etat$)}.
\label{lambdaalphaetat}
\EN
However, if the resulting magnetic field really were to grow like
$e^{\lambda t}$, the effective values of $\alpha$ and $\etat$ would
be modified and would no longer be constant.
By applying \Eqs{alpha_omega_osc}{etat_omega_osc} for a range of values of $\lambda$
for which $1+\lambda\tau>0$ we find that $\alpha$ and $\etat$ become
\EQ
\alpha(\lambda)=\alpha_0 A_\alpha
\frac{1+\lambda \tau_{\alpha}}
{(1+\lambda \tau_{\alpha})^2+\omega_{\alpha}^2\tau_{\alpha}^2}
\label{alpha_lambda}
\EN
and
\EQ
\etat(\lambda)=\eta_{t0} A_{\etat}
\frac{1+\lambda \tau_{\etat}}
{(1+\lambda \tau_{\etat})^2+\omega_{\etat}^2\tau_{\etat}^2}
\label{etat_lambda}
\EN
respectively.
In these equations the occurrence of the terms $\omega_i\tau_i$ for
$i=\alpha$ or $\etat$ is qualitatively new compared with earlier
expectations based on the $\tau$ approximation; see \Sec{TauApprox}.
Note that the relaxation times $\tau_i$ and oscillation frequency $\omega_i$
from \Eq{predic} are in general different for $\alpha$ and $\etat$;
see \Tab{Tsummary}.

A more direct way of calculating $\alpha(\lambda)$ and $\etat(\lambda)$
is by using exponentially growing or decaying test functions proportional
to $e^{st}$, provided that $1+s\tau>0$, which sets the maximal decay rate for which
equations (\ref{alpha_lambda}) and (\ref{etat_lambda}) are meaningful.
The existence of a maximal decay rate is interesting: in such a system the fluctuating
fields survive long enough to preserve the mean field.  Clearly then, solutions of 
\Eq{lambdaalphaetat},
\EQ
\lambda=\alpha(\lambda) k-[\eta+\etat(\lambda)]k^2
\EN
are required for
self-consistent systems (be they dynamos or decaying mean fields).

\subsection{Wavenumber dependence}

In the work of Brandenburg et al.\ (2008a), which led to this paper,
similar methods
were used to find the dependences of $\alpha$ and $\etat$ on the
wavenumber $k$ of the mean magnetic field.
In that paper, it was shown for the Roberts flow that under FOSA we have
\EQ
\tilde\alpha(k)=\frac{\alpha_0}{1+(a_\alpha k/\kf)^2},\quad
\tilde\etat(k)=\frac{\etatz}{1+(a_\eta k/\kf)^2}, \label{alphaofk}
\EN
where $a_\alpha=a_\eta=1$.
They found that this result is also a good approximation to turbulent flows,
but then $a_\alpha$ and $a_\eta$ were treated as fit parameters that are
of order unity.
While that work noted that memory effects should be expected, they were 
not treated.
\EEq{alphaofk} can be directly compared to \Eq{alpha_lambda}
with the growth rate $\lambda$ set to $0$ which recaptures the test-field
method as used in Brandenburg et al.\ (2008a).
This might suggest that $\cos \omega_0t $
is related to the advection term in \Eq{predic},
so one might expect that $\omega_0 \sim k \urms$.
For $\omega_0=\St_i k \urms$ then, the formulae from \Eq{alpha_lambda} and
\Eq{alphaofk} match exactly, and by capturing the dependency of $\alpha$
and $\etat$ on past times, we are perforce treating the problem as also
non-local in space.
One might therefore be tempted to suggest that the combined dependence on
$\omega$ and $k$ could be of the form
\EQ
\tilde\alpha(k,\omega)=\alpha_0 A_\alpha
\frac{1-\ii\omega \tau_{\alpha}}
{(1-\ii\omega \tau_{\alpha})^2+(a_\alpha k/\kf)^2},
\label{alpha_omegak}
\EN
and
\EQ
\tilde\etat(k,\omega)=\eta_{t0} A_{\etat}
\frac{1-\ii\omega \tau_{\etat}}
{(1-\ii\omega \tau_{\etat})^2+(a_\eta k/\kf)^2},
\label{etat_omegak}
\EN
However, although such a formula is indeed obeyed in the two special cases
$\omega=0$ (Brandenburg et al.\ 2008a) and $k=k_1$ (present work),
some preliminary work suggests that this equation is not valid in
general, and that a multiplicative relation of the form
$\tilde\alpha(k,\omega)=\tilde\alpha(k)\tilde\alpha(\omega)$ and
$\tilde\etat(k,\omega)=\tilde\etat(k)\tilde\etat(\omega)$
might be more accurate.

\subsection{Linear time dependence}

After our paper appeared as preprint (arXiv:0811.2561v1),
Hughes \& Proctor (2009) pointed out an inconsistency in the
turbulent magnetic diffusivity tensor
when allowing mean fields with a linear time dependence.
They attributed this to the occurrence of a new contribution
to the magnetic diffusivity.
In the following, we explain that their result is a natural consequence
of using \Eqs{alpha_omega_osc}{etat_omega_osc}, as advocated in our paper.

The time dependence of the mean field in the paper by
Hughes \& Proctor (2009) is given by
\EQ
\meanBB(t)=\BB_0+\CC_0t,
\EN
with constants $\BB_0$ and $\CC_0$.
If we convolve this mean field with the kernels $\hat{\alpha}$
and $\hat{\eta}$, corresponding to the $\tau$ approximation
(i.e.\ proportional to $e^{-t/\tau_\alpha}$ and $e^{-t/\tau_\eta}$,
respectively), we find the $\meanEMF$ to be
\EQ
\meanEMF(t)=(\alpha_0-\etatz k)(\BB_0+\CC_0 t)
-(\tau_\alpha\alpha_0-\tau_\eta\etatz k)\CC_0.
\label{H&P1st}
\EN
This formulation matches the form of Equation (25) of Hughes \& Proctor (2009),
where their $\Gamma$ is given by $-(\tau_\alpha\alpha_0-\tau_\eta\etatz k)$.
Re-expressing \Eq{H&P1st} in terms of $\meanBB(t)$ and
$\partial\meanBB/\partial t$, as well as their curls, proportional
$\meanJJ(t)$ and $\partial\meanJJ/\partial t$,
we can write \Eq{H&P1st} in the form
\EQ
\meanEMF=\alpha_0\meanBB-\etatz\mu_0\meanJJ
+\Gamma_\alpha{\partial\meanBB\over\partial t}
-\Gamma_\eta\mu_0{\partial\meanJJ\over\partial t},
\label{H&P1stGamma}
\EN
where $\Gamma_\alpha=-\alpha_0\tau_\alpha$
and $\Gamma_\eta=-\etatz\tau_\eta$ quantify
additional contributions to the mean electromotive force.
In the more general case where $\omega_\alpha$ and $\omega_\eta$ are
different from zero, we have
\EQ
\Gamma_\alpha=-\alpha_0\tau_\alpha
{1-\omega_\alpha^2\tau_\alpha^2\over(1+\omega_\alpha^2\tau_\alpha^2)^2},\quad
\Gamma_\eta=-\eta_0\tau_\eta
{1-\omega_\eta^2\tau_\eta^2\over(1+\omega_\eta^2\tau_\eta^2)^2}.
\EN

We recall that the formulation in \Eq{H&P1stGamma} only applies to the
special case of variations of the mean field that are linear in time.
More generally, we have
\EQ
\meanEMF = \sum_{n=0}^\infty (-1)^n \left(
\alpha^{(n)}\frac{\partial^n \meanBB}{\partial t^n} 
-\eta^{(n)}\mu_0\frac{\partial^n \meanJJ}{\partial t^n}\right),
\EN
where
\EQ
\alpha^{(n)}=\int_0^{\infty} \hat{\alpha}(t)\,t^n\,\dd t,\quad
\etat^{(n)}=\int_0^{\infty} \hat{\eta}(t)\,t^n\,\dd t.
\EN
These moments are related to the derivatives of $\tilde\alpha(\omega)$
and $\tilde\etat(\omega)$ at $\omega=0$ with
\EQ
\alpha^{(n)}=(-\ii)^n\,\left.{\dd^n\tilde{\alpha}\over\dd\omega^n}\right|_0,\quad
\etat^{(n)}=(-\ii)^n\,\left.{\dd^n\tilde{\eta}\over\dd\omega^n}\right|_0,
\EN
where the subscripts 0 indicate that the derivatives are to be evaluated
at $\omega=0$.
Note, in particular, that
$\Gamma_\alpha=-\alpha^{(1)}=\mbox{Im}(\dd\tilde\alpha/\dd\omega)_0$.

\begin{figure}[t!]
\centering\includegraphics[width=\columnwidth]{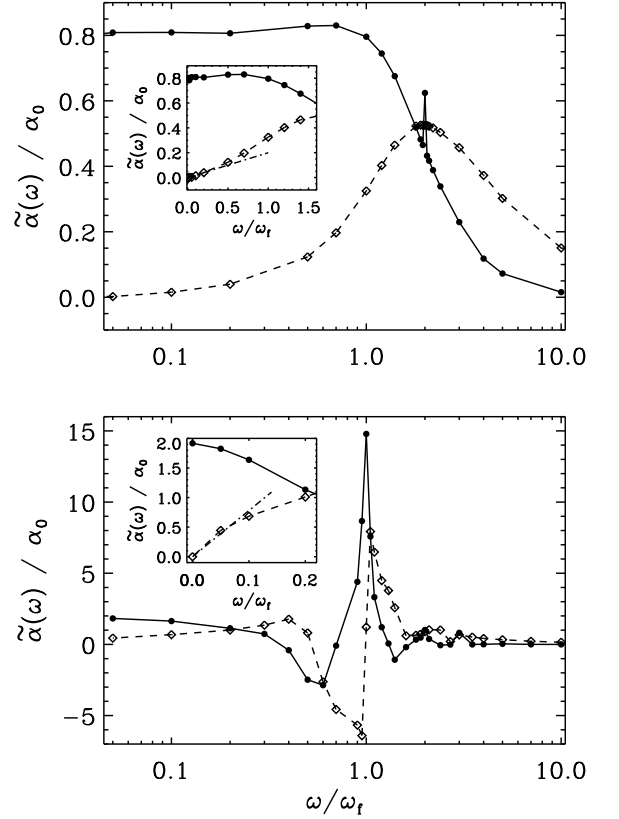}\caption{
Real and imaginary parts of $\tilde\alpha(\omega)$ for $k=0$
using the Otani (1993) MW+ flow with $\Rm=1$ (upper panel) and
$\Rm=100$ (lower panel).
The normalization is chosen to be $\alpha_0=u_0$.
The insets show the scaling of $\mbox{Im}\,\tilde\omega$ near the origin
with slopes 0.2 and 7.8 for the upper and lower panels, in agreement
with the results of Hughes \& Proctor (2009).
}\label{Otani_Rm10_om_po32}\end{figure}

Hughes \& Proctor (2009) have computed values of $\Gamma_\alpha$ for
$\Rm$ between 1 and 100 using
a particular form of the modulated wave flow of Otani (1993),
referred to as MW+ flow, which is given by \Eq{Rob1a}
with $\kf=k_0$ and
\EQ
\psi(x,y,t)={2u_0\over k_0}
\left(\cos^2\!\omega_{\rm f} t\cos k_0x
-\sin^2\!\omega_{\rm f} t\cos k_0y\right),
\EN
where $\omega_{\rm f}=u_0 k_0$ has been chosen.

In order to substantiate our interpretation of their results
we have computed $\tilde\alpha(\omega)$ for their case with $k=0$.
The result for the Fourier-transformed kernel is shown in
\Fig{Otani_Rm10_om_po32} for the Otani MW+ flow with $\Rm=1$ and 100.
Compared with \Fig{o0_dependence}, there are additional features
related to resonances with the frequency $\omega_{\rm f}$ of the Otani flow.
Such features cannot be explained with our simple fit formula.
This means that higher order terms will become important in those cases
where the variation of the mean magnetic field is more complicated than
just a linear increase.

The value of $\Gamma_\alpha$ can readily be read off as the slope
of the graph of $\mbox{Im}\,\tilde\omega$ near the origin.
Our results agree with those of Hughes \& Proctor (2009), as is shown
in the insets of \Fig{Otani_Rm10_om_po32}.
We note, however, there are no good reasons to associate the $\Gamma_\alpha$
term with a correction to turbulent diffusion alone.
Instead, there are corrections both to $\alpha$ and to $\etat$ once
the mean magnetic field shows strong time dependence.

In this connection it is important to emphasize that these complications
are mainly a consequence of the particular time dependence inherent
to the Otani flow and are not typical of turbulence, as seen before.
For $\Rm=100$ there is a distinct spike at $\omega=\omega_{\rm f}$,
while for $\Rm=1$ there is a smaller spike at $\omega=2\omega_{\rm f}$;
see \Fig{Otani_Rm10_om_po32}.
We hypothesize that these spikes
are associated with the periodicity of the Otani flow.
Similar behavior is known to occur for the Galloway \& Proctor (1992) flow
(Courvoisier et al.\ 2006), and is connected with the infinite
correlation time of a flow with sinusoidal time dependence
(R\"adler \& Brandenburg 2009).

\section{Conclusions}
\label{Conclusions}

Naive application of the values of $\alpha$ and $\etat$ to time-dependent
problems can lead to errors.
This is because the turbulent transport coefficients are in general
frequency-dependent, due to memory effects.
So, for each frequency and for each growth or decay rate
(corresponding to imaginary frequencies) the transport coefficients
need to be determined separately.
The full frequency dependence can then be used to calculate response
functions via Fourier transformation.
The result can then be used to determine the response to general time
dependences, including, for example, oscillatory growth.

The response function formalism 
shows that one needs to know the past time history of the mean
fields to compute turbulent transport correctly.
This is not new, but what is new is the fact that the departures from
the instantaneous approximation can be quite substantial for flows
such as the Roberts flow.
For isotropic turbulence, on the other hand, the effects tend to
be less dramatic and simple fit formulae with an exponential decay
and an oscillatory part can be reasonably accurate.

The presence of an oscillatory part in the response function
proportional to $\cos\omega_0 t$ leads to a sign reversal
of $\alpha$ and $\etat$.
Hori \& Yoshida (2008) associate this with the ``over-twisting'' in
illustrations of Parker's $\Omega$ loops.
In their picture, rising flux tubes may twist by more than $90^\circ$.
This interpretation clarifies the naive expectation that $\alpha$ may
change the sign when the Coriolis force becomes important.
In fact, as our work now shows, such an effect would only occur
if the mean magnetic field varies like a $\delta$ function in time
or if it shows other rapid variations.
Conversely, for mean fields varying slowly in time the net $\alpha$
would not change the sign, although some past times are weighted negatively.

In the present work, we have only looked at one type of memory effect,
where the typical timescales in the integral kernel are comparable to
the dynamical timescales of the turbulence.
There is yet another type of memory effect that can occur on a resistive
timescale, namely the one associated with magnetic helicity conservation.
As explained in the appendix of Blackman \& Brandenburg (2002), this is
a purely nonlinear effect such that the relevant time scale becomes very
long only when the magnetic field is strong.
Obviously, this effect is not captured by our kinematic approach, nor
would it be relevant in connection with the calculation of growth rates
of the dynamo.

The approach presented here may be useful for calculating memory
effects of turbulent transport coefficients over a range of other
related problems.
Particularly important might be the question of the damping of
acoustic waves by turbulent viscosity in the Sun, for example
(Stix et al.\ 1993).
Such damping would lead to line broadening of the acoustic frequencies.
The present work has demonstrated that such quantities can only be
useful if one has a good idea of its frequency dependence relative
to the frequency at which the turbulent viscosity is determined
and the frequency at which it is to be applied.

Our approach could also be useful in cases where the turbulence
itself is time dependent.
This would be relevant for modeling convection in pulsating stars.
Such systems are currently being treated with time-dependent
mixing length theory (Gough 1977).
It would seem appropriate to adopt integral kernels also in that case.
However, now there would be two frequencies to be considered: the
frequency at which the turbulence varies and the frequency
at which the mean field varies.
Another problem is that the test-field method has only been used
and tested in connection with magnetic and passive scalar diffusion
problems, and has not yet been developed for calculating the components
of the turbulent viscosity tensor.
This would indeed be one of the outstanding problems in this field.

\acknowledgements

We thank the referee for suggesting many improvements to the paper
and for presenting us with the calculation that is now reproduced in
\App{appB}.
We acknowledge Matthias Rheinhardt for making useful suggestions.
The computations have been carried out on the
National Supercomputer Centre in Link\"oping and the Center for
Parallel Computers at the Royal Institute of Technology in Sweden.
This work was supported in part by the Swedish Research Council,
grant 621-2007-4064, and the European Research Council under the
AstroDyn Research Project 227952.

\appendix
\section{Convolution for monochromatic variations}
\label{convolutiont}

The purpose of this appendix is to show that for monochromatic
signals a convolution corresponds to a multiplication in real space.
Consider \Eq{convolutiontEq} for a monochromatic function
\EQ
\meanGG(t)=\meanGG_\omega(t)\equiv\GG_0\cos\omega t, \label{convolutiontEq}
\EN
where $\omega$ is a constant.
Inserting this into \Eq{convolutiont1} yields
\EQ
\meanFFF_\omega(t)
=-\int_{-\infty}^\infty\hat\kappat(t-t')\GG_0\cos\omega t'\,\dd t'
=-\GG_0\mbox{Re}\int_{-\infty}^\infty\hat\kappat(t-t') e^{-\ii\omega t'}\,\dd t'
=-\GG_0\mbox{Re}\,e^{-\ii\omega t}\int_{-\infty}^\infty\hat\kappat(t-t') e^{\ii\omega(t-t')}\,\dd t'.
\EN
By using a change of variables one sees that the integral is just the
Fourier transform of $\hat\kappat(t)$.
Thus, we arrive at
\EQ
\meanFFF_\omega(t)
=-\GG_0\mbox{Re}\left[e^{-\ii\omega t}\tilde\kappat(\omega)\right].
\label{Ftilde_omega}
\EN
The real part of $\tilde\kappat$ shows therefore a modulation with
$\cos\omega t$ and the imaginary part with $\sin\omega t$.
By projecting against these two functions separately, we can
determine the real and imaginary parts of $\tilde\kappa(\omega)$.
Thus, the complex function $\tilde\kappa(\omega)$ can be obtained
from $\meanFFF_\omega(t)$ as
\EQ
\tilde\kappat(\omega)=-2G_0^{-1}\left\langle e^{\ii\omega t}
\meanFFF_\omega(t)\right\rangle_t\,, \label{appAkt}
\EN
which is the result stated in \Eq{mean_eiot_F}.
The factor 2 stems from the fact that the average values of
$\cos^2\omega t$ and $\sin^2\omega t$ are 1/2.
This procedure can be trivially extended to tensorial relationships;
cf.\ \Eq{mean_est_E}.

It is interesting to write \Eq{Ftilde_omega} by expressing
$\tilde\kappat(\omega)$ in terms of its modulus and its phase,
$|\tilde\kappat|\exp\ii\phi_\kappa$, so we have
\EQ
\meanFFF_\omega(t)
=-|\tilde\kappa|\,\meanG(t-\Delta t),\quad
\mbox{where $\Delta t=\phi_\kappa(\omega)/\omega$},
\label{TimeDelay}
\EN
showing that memory effects change not only the amplitude of the effective
transport coefficient, but they also lead to a time lag such that, for
a given frequency, the mean flux is proportional to the mean fields at
a certain later time.

\section{Roberts flow with oscillatory mean concentration gradient}
\label{appB}

As was generously pointed out by the referee, in the special case of
a Roberts flow, Eqs.~(\ref{Rob1a}) and (\ref{Rob2a}), with a mean
concentration $\meanC=zG_0\cos\omega t$, we can solve the problem
analytically.
In this case, Equation~(\ref{dcdt}) becomes
\EQ
{\partial c\over\partial t}=-u_z G_0 \cos\omega t
-\nab\cdot(\uu c-\overline{\uu c})+\kappa\nabla^2c.
\EN
In a first step we employ FOSA and neglect
$\nab\cdot(\uu c-\overline{\uu c})$, so the above reduces to
\EQ
{\partial c\over\partial t}-\kappa\nabla^2c
=-\sqrt{2} u_0 G_0 \cos k_0 x \cos k_0 y \cos\omega t.
\label{modeqappb}
\EN
This has as a solution
\EQ
c(x,y,t)=-\frac{\sqrt{2} u_0 G_0}{\left(\omega^2+(\kappa\kf^2)^2)\right)^{1/2}}
\cos k_0 x\cos k_0 y\cos(\omega t-\phi),
\label{cappb1}
\EN
where
\EQ
\cos\phi=\frac{\kappa\kf^2}{\left(\omega^2+(\kappa\kf^2)^2\right)^{1/2}},
\quad
\sin\phi=\frac{\omega}{\left(\omega^2+(\kappa\kf^2)^2\right)^{1/2}}.
\EN
Note that $\nab \cdot (\uu c -\overline{\uu c})=0$ and so this particular
solution is also valid beyond FOSA.
We obtain then
\EQ
\meanFFFF^{\omega}(t)=-\frac{u_0^2 G_0}{2 \left(\omega^2+(\kappa\kf^2)^2\right)^{1/2}} \cos(\omega t-\phi)\zzz.
\EN
We can now find the Fourier-transformed kernel through \Eq{appAkt}:
\EQ
\tilde\kappat(\omega)=-2G_0^{-1}\left\langle e^{\ii\omega t}
\meanFFF_\omega(t)\right\rangle_t=\frac{u_0^2}{2\left(\omega^2+(\kappa\kf^2)^2\right)^{1/2}}\left(\frac{\kappa\kf^2+i \omega}{\left(\omega^2+(\kappa\kf^2)^2\right)^{1/2}}\right)=\frac{\tau_k u_0^2}{2}\left(\frac{1+i \omega \tau_k}{1+\omega^2 \tau_k^2}\right)=\frac{\tau_k u_0^2}{2}\left(\frac{1}{1-\ii\omega \tau_k}\right),
\label{appbeqfinal}
\EN
where we have defined $\tau_k^{-1}=\kappa \kf^2$.
Equations~(\ref{Rob1a}) and (\ref{Rob2a}) imply that
$\overline{u_z^2}=u_0^2/2$, and \Sec{PSDiff} argues that $\tau=\tau_k$.
Accordingly, \Eq{appbeqfinal} reduces to \Eq{kappa_omega}.

\section{Simplified one-dimensional model}
\label{1D}

\begin{figure}[ht!]\begin{center}
\includegraphics[width=.5\columnwidth]{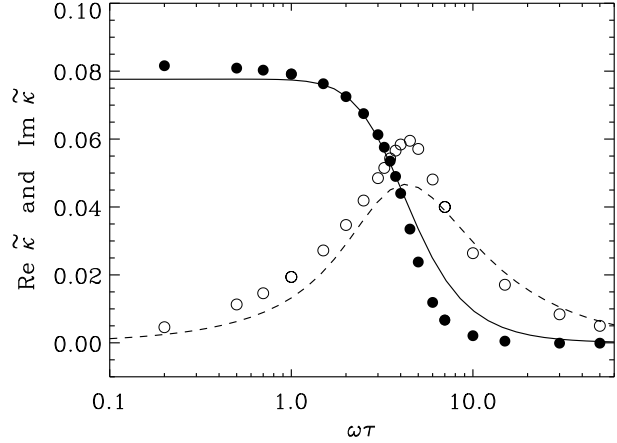}
\end{center}\caption[]{
Real (solid circles) and imaginary (open circles) components
of $\kappa_{\rm t}$ for $k_z=5$ and $\uu=u_0\zzz \cos x$ (see \App{1D}),
using as fit parameters $A_\kappa=0.105$, $\tau_\alpha=0.33$, and
$\omega_\kappa=1.8$.
}\label{dcdt_complex_k5}\end{figure}

A simple system that defies result \eq{kappa_omega} of the $\tau$ approximation is one
with a passive scalar whose concentration varies sinusoidally along $z$ with $k_z\neq0$ and
a steady flow $\uu=(0,0,u)$, such that $u=u(x)=u_0\cos k_0x$,
so $\nab\cdot\uu=0$.  The equation for the small-scale concentration then is
\EQ
{\partial c\over\partial t}=-\nab\cdot(\uu\meanC+\uu c-\overline{\uu c})
+\kappa\nabla^2c,
\EN
which becomes
\EQ
{\partial c\over\partial t}=-u\meanG
-u{\partial c\over\partial z}+\overline{u{\partial c\over\partial z}}
+\kappa\nabla^2c,
\EN
and in turn
\EQ
{\partial c\over\partial t}=
-u{\partial c\over\partial z}+\overline{u{\partial c\over\partial z}}
+\kappa\nabla^2c-u(x)\meanG(t,z).
\EN
This system is linear, inhomogeneous, with variable coefficients.
We note that $\meanG(t,z)=G(t)e^{ik_zz}$, impose
\EQ
G(t)=G_0 \cos \omega t,
\EN
assume that
\EQ
c=\tilde{c}(t,x)e^{ik_zz}+\mbox{c.c.}
\EN
and treat as the system as a time-dependent problem
with complex $\tilde{c}(t,x)$:
\EQ
{\partial\tilde{c}\over\partial t}=-\left[\ii k_z u(x)+\kappa k_z^2\right]\tilde{c}
+\kappa{\partial^2\tilde{c}\over\partial x^2}
+\ii k_z\overline{u\tilde{c}}-u_0 G_0 \cos k_0 x \cos \omega t.
\label{dtildecdt}
\EN
\Eq{dtildecdt} is the equivalent equation to \Eq{modeqappb} (and reduces to that equation
when $k_z=0$).  The Fourier-transformed kernel can be calculated similar to \App{appB}, and
in \Fig{dcdt_complex_k5} we present a numerical solution for the Fourier-transformed
kernel for $k_z/k_1=5$ and $u_0/\kappa k_z=5$.

\end{document}